\input phyzzx
\hsize=38pc
\font\sf=cmss10                    
\input BoxedEPS
\SetRokickiEPSFSpecial 
\HideDisplacementBoxes 
\def\Figure#1#2{\midinsert
$$\BoxedEPSF{#1}$$
\noindent {\sf #2}
\endinsert}

\def\bbbc{{\mathchoice {\setbox0=\hbox{$\displaystyle\rm C$}\hbox{\hbox
to0pt{\kern0.4\wd0\vrule height0.9\ht0\hss}\box0}}
{\setbox0=\hbox{$\textstyle\rm C$}\hbox{\hbox
to0pt{\kern0.4\wd0\vrule height0.9\ht0\hss}\box0}}
{\setbox0=\hbox{$\scriptstyle\rm C$}\hbox{\hbox
to0pt{\kern0.4\wd0\vrule height0.9\ht0\hss}\box0}}
{\setbox0=\hbox{$\scriptscriptstyle\rm C$}\hbox{\hbox
to0pt{\kern0.4\wd0\vrule height0.9\ht0\hss}\box0}}}}

\catcode`\@=11 
\def\NEWrefmark#1{\step@ver{{\;#1}}}
\catcode`\@=12 

\def\footstrut{\baselineskip 12pt}
\def\square{\kern1pt\vbox{\hrule height 1.2pt\hbox{\vrule width 1.2pt\hskip 3pt
   \vbox{\vskip 6pt}\hskip 3pt\vrule width 0.6pt}\hrule height 0.6pt}\kern1pt}

\def\bra#1{\langle #1 |}
\def\ket#1{| #1 \rangle}
\def\vev#1{\langle #1 \u
ngle}

\def\ov{{\overline}}

\def\A{{\cal A}}
\def\B{{\cal B}}
\def\C{{\cal C}}

\def\H{{\cal H}}

\def\L{{\cal L}}
\def\M{{\cal M}}

\def\O{{\cal O}}
\def\P{{\cal P}}

\def\V{{\cal V}}

\def\e{{\epsilon}}

\def\ov{\overline}

\def\B{{\cal B}}

\def\P{{\cal P}}
\def\V{{\cal V}}
\def\O{{\cal O}}

\singlespace

\def\define#1#2\par{\def#1{\Ref#1{#2}\edef#1{\noexpand\refmark{#1}}}}
\def\con#1#2\noc{\let\?=\Ref\let\<=\refmark\let\Ref=\REFS
         \let\refmark=\undefined#1\let\Ref=\REFSCON#2
         \let\Ref=\?\let\refmark=\<\refsend}

\let\refmark=\NEWrefmark

\define\barton{B. Zwiebach, {\it Quantum open string theory
with manifest closed string factorization}, Phys. Lett. {\bf B256} (1991) 22}

\define\zwiebachos{B. Zwiebach, {\it Interpolating string field theories},
Mod. Phys. Lett. {\bf A7} (1992) 1079, hep-th/9202015.}

\define\senzwiebachtwo{A.~Sen and B.~Zwiebach, {\it Quantum background
independence of closed string field theory}  Nucl. Phys.{\bf
B423} (1994) 580, hep-th/9311009.}

\define\senzwiebachnew{A.~Sen and B.~Zwiebach, {\it Background
independent algebraic structures in closed string field theory},
Comm. Math. Phys. {\bf 177} (1996) 305, hep-th/9408053.}

\define\zwiebachlong{B. Zwiebach, `Closed string field theory: Quantum
action and the Batalin-Vilkovisky master equation', Nucl. Phys {\bf B390}
(1993) 33, hep-th/9205075.}

\define\wittenosft{ E. Witten,  {\it Noncommutative geometry and 
string field theory}, Nucl. Phys. {\bf B268} (1986) 253.}

\define\freedman{D. Z. Freedman, S. B. Giddings, J. A. Shapiro,
and C. B. Thorn,{\it  The nonplanar one-loop amplitude in
Witten's string field theory}, Nucl. Phys. {\bf  B287} (1987) 61}

\define\zwiebach{ B. Zwiebach, {\it How covariant closed string 
theory solves a minimal area problem}, Comm. Math. Phys. {\bf 136} (1991) 83;
 {\it Consistency of closed string
polyhedra from minimal area}, Phys. Lett. {\bf B241} (1990) 343}

\define\sonodazw{ H. Sonoda and B. Zwiebach,  {\it Closed string field theory
loops with symmetric factorizable quadratic differentials}, 
Nucl. Phys. B331 (1990) 592.}

\define\zwiebachqcs{ B. Zwiebach,  {\it Quantum closed strings from minimal
area},  Mod. Phys. Lett. A5 (1990) 2753.}

\define\thornpr{C. B. Thorn, {\it String field theory}, Phys. Rep. 
{\bf 174} (1989) 1.}

\define\bogojevic{ A. R. Bogojevic,  {\it BRST Invariance of the measure
in string field theory}, Phys. Lett.  {\bf B198} (1987) 479}

\define\maeno{ M. Maeno,  {\it Canonical quantization of Witten's string field
theory using midpoint light cone time} Phys. Rev. {\bf D43} (1991) 4006}

\define\morris{ T. R. Morris, Nucl. Phys. {\bf B297} (1988) 141;\hfill\break
J. L. Ma\~nes, Nucl. Phys. {\bf B303} (1988) 305;\hfill\break
M. Maeno, Phys. Lett. {\bf B216} (1989) 81;\hfill\break
R. Potting and C. Taylor, Nucl. Phys. {\bf B316} (1989) 59;\hfill\break
T. Kugo, Prog. Theor. Phys. {\bf 78} (1987) 690;\hfill\break
S. J. Sin, Nucl. Phys. {\bf B306} (1988) 282; B313 (1989) 165;\hfill\break
W. Siegel and B. Zwiebach, Nucl. Phys. {\bf B282} (1987) 125;\hfill\break
P. Goddard, J. Goldstone, C. Rebbi and C. B. Thorn, Nucl. Phys.
{\bf B56} (1973) 109}

\define\saitoh{ Y. Saitoh and Y. Tanii, Nucl. Phys. {\bf B325} (1989) 161;
Nucl. Phys. {\bf B331} (1990) 744.}

\define\kikkawa{ K. Kikkawa and S. Sawada,  {\it Cancellation of the
Lorentz anomaly of string field theory in light-cone gauge},
Nucl. Phys. {\bf B335} (1990) 677.}

\define\hatanojiri{H. Hata and M. M. Nojiri, {\it New symmetry in
covariant open-string field theory}, Phys. Rev {\bf D36} (1987) 1193} 

\define\callan{ C. G. Callan, D. Friedan, E. Martinec and M. Perry,
Nucl. Phys. B262 (1985) 593;\hfill\break
A. Sen, Phys. Rev. D32 (1985) 2102; Phys. Rev. Lett. 55 (1985) 1846}

\define\schwarz{A. Schwarz, {\it Grassmannians and string theory}, hep-th/9610122.}

\define\shapiro{ J. A. Shapiro and C. B. Thorn, {\it Closed string-open 
string transitions and Witten's string field theory}, Phys. Lett
{\bf B194} (1987) 43}

\define\alvarez{L. Alvarez-Gaume, C. Gomez, G. Moore and C. Vafa, Nucl. 
Phys. {\bf B303} (1988) 455}

\define\polch{J. Polchinski, {\it Tasi Lectures on D- branes}, hep-th/9611050.}

\define\astashkevich{A. Astashkevich and A. Belopolsky, {\it String center of
mass operator and its effect on BRST cohomology}, to appear in
Comm. Math. Phys. hep-th/9511111. }

\define\belopolskyzw{A. Belopolsky and B. Zwiebach, {\it Who changes the
string coupling ? } Nucl. Phys. {\bf B472} (1996) 109, hep-th/9511077.}

\define\zwiebachhms{B. Zwiebach, {\it New moduli spaces from background
independence consistency conditions},  Nucl.~Phys.
{\bf B480} (1996) 507,  hep-th/9605075 } 

\define\stasheff{ J.D. Stasheff, {\it On the homotopy associativity of
$H$-spaces, I.}, Trans. Amer. Math. Soc. {\bf 108}, 275 (1963);
{\it On the homotopy associativity of $H$-spaces, II.},
Trans. Amer. Math. Soc. {\bf 108}, 293 (1963). }

\define\zwiebachnc{B. Zwiebach, {\it String field theory around non-conformal
backgrounds} ,  Nucl.~Phys. {\bf B480} (1996) 541, hep-th/9606153 }

\define\zwiebachopen{B. Zwiebach,  {\it A proof that Witten's open string theory gives a single cover  of moduli space}, Comm. in Math. Phys. {\bf 142} (1991) 193.}

\define\zwiebachopentwo{B. Zwiebach, {\it 
Minimal area problems and quantum open strings}
, B.~Zwiebach, Comm. in Math. Phys.{\bf 141} (1991) 557.}

\define\hatazwiebach{H. Hata and B. Zwiebach, 
{\it Developing the covariant Batalin-Vilkovisky
approach to string theory} , Ann. Phys. {\bf 229} (1994) 177, hep-th/9301097.}

\define\gaberdielzwiebach{M. Gaberdiel and B. Zwiebach, 
{\it Tensor constructions of Open String
Theories I: Foundations}, hep-th/9705038.}

\define\bergmanzwiebach{O. Bergman and B. Zwiebach, {\it The dilaton theorem and
closed string backgrounds}, Nucl. Phys. {\bf  B441} (1995) 76, hep-th/9411047}

\define\rahmanzwiebach{S. Rahman and B. Zwiebach, {\it Vacuum vertices and the
ghost-dilaton} Nucl. Phys. {\bf B471} (1996) 233, hep-th/  }

\define\sagnotti{M. Douglas and B. Grinstein,
 {\it Dilaton tadpole for the open bosonic string}
Phys. Lett. {\B183} (1987) 52; \hfill\break
N. Marcus and A. Sagnotti, {\it Group theory from quarks at the ends
of strings}, Phys. Lett. {\bf B188} (1987) 58.}

\define\callan{C. Callan, C. Lovelace, C. Nappi and S. Yost, {\it Adding holes 
and crosscaps to the superstring}, Nucl. Phys. {\bf 293} (1987) 83.}

\def\footstrut{\baselineskip 12pt}
\def\square{\kern1pt\vbox{\hrule height 1.2pt\hbox{\vrule width 1.2pt\hskip 3pt
   \vbox{\vskip 6pt}\hskip 3pt\vrule width 0.6pt}\hrule height 0.6pt}\kern1pt}

\def\bra#1{\langle #1 |}
\def\ket#1{| #1 \rangle}
\def\vev#1{\langle #1 \rangle}

\singlespace
{}~ \hfill \vbox{\hbox{MIT-CTP-2644}
\hbox{HUTP-97/A025}
\hbox{hep-th/9705241}\hbox{
} }\break
\title{ORIENTED OPEN-CLOSED STRING THEORY REVISITED}
\author{Barton Zwiebach \foot{E-mail address: zwiebach@irene.mit.edu
\hfill\break Supported in part by D.O.E.
contract DE-FC02-94ER40818, and a fellowship of the John Simon Guggenheim
Memorial Foundation.}}
\address{Lyman Laboratory of Physics,\break
Harvard University \break
Cambridge, Massachusetts 02138, U.S.A.}

\abstract 
{String theory on  D-brane backgrounds is open-closed
string theory. Given the relevance of this fact, we give 
details and elaborate upon our earlier construction of oriented 
open-closed string field theory. 
In order to incorporate explicitly closed strings, the classical
sector of this theory is  open strings  with a homotopy associative $A_\infty$ algebraic
structure. 
We build a
suitable Batalin-Vilkovisky   
algebra on  moduli spaces of bordered Riemann surfaces, 
the construction of which involves a few subtleties arising from the open string
punctures and cyclicity conditions. 
All vertices coupling open and closed strings through disks are described explicitly.  
Subalgebras of the algebra of surfaces with boundaries are used to
discuss symmetries of classical open string theory induced
by the closed string sector, and to write classical open string field 
theory on general closed string backgrounds. We give a preliminary analysis
of the ghost-dilaton theorem. }
\endpage

\singlespace

\chapter{Introduction and Summary} 

Recent developments  indicate that the distinction between 
theories of open and closed strings and theories of closed strings 
is not fundamental. Theories that are formulated as pure closed string
theories on simple backgrounds may 
show open string sectors on
D-brane backgrounds (for a review, see [\polch]). 
In  such backgrounds open-closed string field theory
gives a complete description of the perturbative physics.
It is therefore of interest to have a good understanding of
open-closed string field theory.
The elegant open string field theory 
of Witten [\wittenosft] is formulated without including
an explicit closed string sector. The price for this simplicity is 
lack of manifest factorization in the closed string 
channels [\freedman,\shapiro]. 
It is now clear that it is generally useful
to have an explicit closed string sector. 
A covariant open-closed string field theory 
achieving  this was sketched in Ref.[\barton]. In 
this theory the Batalin Vilkovisky (BV) master equation is 
satisfied manifestly. One of the  purposes of
the present paper is to give the detailed construction promised 
in Ref.[\barton]. Much of that  was actually written in 1991. 
In completing this paper now we have taken the
oportunity to develop and explain some further 
aspects of open-closed theory.   Since understanding the
conceptual unity underlying closed and open-closed string
theory is important, 
we will show  that open-closed string theory and closed string theory
are simply two  different concrete realizations of the same
basic geometrical and algebraic structures.

The  structures include a set of spaces $\P$
where one can define an antibracket $\{ \, , \, \}$ 
and a delta operation $\Delta$
satisfying the axioms of the corresponding BV operators. 
From the spaces $\P$ one constructs a space  
$\V$ satisfying the condition $\partial \V + \half \{ \V, \V \} +
\Delta \V =0$. Here $\partial$ is the boundary operator. In addition, one has a 
vector space ${\cal H}$, equipped with an antibracket and 
delta defined on $C({\cal H})$ (functions on ${\cal H})$, and 
an odd quadratic function $Q$ satisfying $\{ Q , Q \}=0$. 
Finally, there is a map $f$ from $\P$
to $C({\cal H})$ inducing a homomorphism between
the respective BV structures, with
the boundary operator $\partial$ realized by the hamiltonian $Q$. 
The  action  is  $S = Q + f(\V)$, and
manifestly satisfies the BV master equation. 

In the case of closed string field theory, where the above structures
were recognized,   $\P$  are moduli spaces
of boundaryless Riemann surfaces having marked points (closed 
string punctures). 
The antibracket operation and delta are realized by twist-sewing
[\senzwiebachtwo, \senzwiebachnew], 
${\cal H}$ is a suitably restricted
 state space of a conformal field theory, and
the function $Q$ is the  BRST hamiltonian. The antibracket on
$C({\cal H})$ arises from a symplectic form in ${\cal H}$. The object $\V$ 
is defined as the formal sum 
of the string vertices $\V_{g,n}$ for all allowed values of $g$ and $n$.
Each string vertex $\V_{g,n}$ represents the region of the moduli space
of Riemann surfaces of genus $g$ and $n$ punctures which cannot
be obtained by sewing of lower dimension string vertices. 

For open-closed string field theory the story, to be elaborated in this
paper, goes as follows. The spaces $\P$ are moduli spaces of bordered
Riemann surfaces having marked points in the interior (closed string
punctures) and marked points on the boundaries (open string punctures).
The antibracket and delta operation are realized by twist-sewing of
closed string punctures and sewing of open string punctures. The vector
space ${\cal H}$ is now of the form ${\cal H} ={\cal H}_o \oplus {\cal H}_c$,
the direct sum of open and  closed string sectors. The BRST operator on
${\cal H}$ is defined by the action of the open BRST operator $Q_o$ on vectors
lying on ${\cal H}_o$ and by the action of the closed BRST operator
$Q_c$ on vectors lying on ${\cal H}_c$. The symplectic structure on
${\cal H}$ arises from separate symplectic structures on ${\cal H}_o$ and
${\cal H}_c$ without mixing of the sectors. The antibracket on $C({\cal H})$
arises from the symplectic structure. The object $\V$ 
is defined as the formal sum 
of the string vertices $\V^{g,n}_{b,m}$ for all allowed values of $g,n,b$ and 
$m$. Each string vertex $\V^{g,n}_{b,m}$ represents a region of the moduli space
of Riemann surfaces of genus $g$, with $n$ closed string punctures, $b$ boundary
components and a total of $m = m_1 + \cdots m_b$ open string punctures
($m_i$ on the boundary component $b_i$). This vertex represents the region
of the moduli space that cannot be obtained by sewing of lower dimensional
vertices.

The above construction was not without some subtleties. Moduli spaces have to
be $Z_2$ graded, and in closed string theory we used the dimension of 
the moduli space (mod 2) as a degree. This was consistent with the fact that
the closed string antibracket, which involves twist sewing and thus adds one real
dimension, is of degree one.  For open strings, the antibracket involves just
sewing, and does not increase dimensionality.  We must therefore find a new
definition of degree.  It turns out that the proper definition is quite natural:
the grade of a moduli space of surfaces $\A^{g,n}_{b,m}$  is the difference between its dimension and the canonical dimension of the moduli space 
$\M^{g,n}_{b,m}$ of surfaces of the same type.  The antibracket can then be verified
to be always of degree one.  In order for the open string antibracket to have the
correct exchange property the moduli spaces in $\P$ must satisfy additional conditions.
Given a boundary component with $m$ open string punctures, these punctures
must be labelled cyclically, and the moduli space must go to itself, up to the sign
factor $(-)^{m-1}$, under a cyclic permutation of the punctures. If there are several
boundary components, these must also be labelled and moduli spaces must
go into themselves up to nontrivial sign factors under the exchange of labels on
the boundary components.

We thus see that open-closed string theory provides a new realization
of  the basic structures found in closed string theory. 
Open-closed string theory is therefore confirming the relevance of 
the structures we have learned about. 
Open-closed string theory is also suggesting the usefulness of
bringing out explicitly each sector of the theory:
it seems better to introduce a closed string sector rather than
having it arise in a singular way. This may be an important lesson,
especially on the light of recent developments that 
show the relevance of the soliton sectors in string theory. 
More sophisticated formulations of string theory
may require some generalization of the above structures, as those discussed in
Refs.[\zwiebachhms,\zwiebachnc] in the context of formulating string
theory around non-conformal backgrounds. 
It is also possible that  $\P$ spaces may satisfy the above axioms without being
moduli  spaces of surfaces.  The possibility of  defining  $\P$ spaces from Grassmannians and using them to build string amplitudes has been studied in Ref.[\schwarz].

This paper begins (section 2) by discussing the moduli spaces of Riemann
surfaces with boundaries and explaining the definition of local coordinates
around open and closed punctures. We discuss the five possible sewing
configurations and learn how to assign a $Z$ degree to moduli spaces. 
We then turn to the definition of the antibracket and introduce the proper cyclic
complex of surfaces.  The discussion of the antibracket is facilitated by
introducing a strictly associative multiplication of moduli spaces (section 2.5).

The string vertex $\V$ comprising the formal sum of all vertices of the
open-closed string theory is introduced in section 3.  We explain how
the recursion relations take the form of a BV type master equation for $\V$.
We give a derivation from first principles of the factors of $\hbar$ and coupling
constant $\kappa$ that accompany each moduli space $\V^{g,n}_{b,m}$
Taking three open string vertex to be at $\hbar^0$ and multiplied by a single
power of $\kappa$, the $\hbar$ and $\kappa$ dependence of all interactions
is naturally fixed  by the BV equation. The closed
string kinetic term  appears at the classical level and all
other interactions at higher orders of $\hbar$.\foot{The same 
seems to hold in  light-cone open-closed string field theory.  
The Lorentz invariance of the classical open string theory and
the classical closed string theory was proven in
refs.[\morris]. The open-closed theory was studied in
[\saitoh] and shown {\it not} to be invariant under the
expected Lorentz transformation.
Kikkawa and Sawada [\kikkawa] have shown
that the non-invariance of the action is actually 
cancelled by the non-invariance of the measure. Introducing
the relevant factors of $\hbar$ in [\kikkawa] the orders
of $\hbar$ of all interactions are fixed once the open string
theory is considered classical. The closed string kinetic
term must be classical, and the open-closed and open-open-closed
interactions appear at order $\hbar^{1/2}$ as they did in
the present work. }
At order $\hbar^{1/2}$, for example, we find the three-closed-string vertex
and all the couplings of one closed
string to $m\geq 0$ open strings through a disk. The string vertex
corresponding to the moduli space of  genus $g$ surfaces with 
$b$ boundary components,
$m$ open strings and $n$ closed strings appears in
order $\hbar^p$, where $p = 2g+b+{1\over 2} n -1$. 
We also isolate useful subfamilies of vertices related by simple
recursion relations. They are: (i) the set of all closed string vertices,
(ii) the set of open string vertices on a disk, (iii)  disks with open strings
and one or zero closed strings, (iv) disks with open strings and two or
less closed strings, (v) disks with all numbers of open and closed strings.

 In section 4 we review the minimal area problem
that defines the string diagrams of open-closed string theory
and extract all string vertices corresponding
to moduli spaces up to real dimension two.  Here we discuss the classical
open string sector of the open-closed theory.  We explain why the strictly
associative version of classical open string theory cannot incorporate
an explicit closed string sector. The algebraic structure of the classical
open string sector is that of an $A_\infty$ algebra [\stasheff] equipped
with additional structure, as explained in [\gaberdielzwiebach].
 In section 5 we discuss the
state spaces of open-closed conformal theories and the symplectic
structures. We give the map from moduli spaces to string functionals,
write the master action and verify  that it satisfies the master equation.

The explicit description of all vertices coupling open and closed 
strings through a disk is the subject of  section 6.  We give some comments
of the ways of describe boundary states $\bra{B}$ both as states in the dual state 
space $\H^*_c$ or as states in $\H_c$. The open-closed vertex is discussed, noting
in particular that  it  induces maps between the BRST cohomologies of the closed 
and open sectors.  These maps take open string cohomology at ghost number
$G$ to closed string cohomology at $G+2$, and closed string cohomology at
ghost number $G$ to open string cohomology at ghost number $G$.
We give the string diagrams for the open-open-closed vertex
and show how it  is generalized to the case of  $M$ open strings coupling to 
a single closed string.  Then we give the string diagram for the open-closed-closed
vertex and use it to generalize to the case of $M$ open strings coupling to two 
closed strings.  Finally, we extend this to the case of $M$ open strings and $N$
closed strings coupling through a disk.

Following the early work of Hata and Nojiri [\hatanojiri], we discuss in section 7 classical
open string  symmetries that arise from the closed string sector. In this
symmetry transformation, the inhomogeneous part of the open string
field shift is along the image, under the cohomology
map mentioned above, of the ghost number one closed string cohomology.  
We elucidate completely the algebra of such transformations.  All these facts
follows easily from analysis of the subalgebras of surfaces involving one or
two closed strings coupling to any number of open strings.  
We then use  the subalgebra of all couplings of open and closed 
strings through disks to construct  in section 8
a gauge invariant classical open string theory describing propagation 
on a nontrivial closed string background.  By the nature of the subalgebras of
surfaces, doing the opposite, {\it i.e.} closed string propagation on open string
backgrounds, seems difficult to achieve.

This paper concludes in section 9 with a brief discussion of the issues that arise
in establishing a ghost dilaton theorem in the open-closed system, and related issues
in background independence.  For the ghost dilaton theorem a few preliminary 
observations are the following.  While the main effect of changing the string
coupling is due to the shift of the closed string field along the ghost-dilaton, an
inhomogeneous shift of the open string field along an unphysical direction
appears to be necessary. This relies on the fact that the image of the ghost-dilaton
under the open-closed cohomology map is a trivial state.  Moreover, a non-vanishing
coupling to a boundary of the ghost-dilaton indicates the presence of a tree level
cosmological term associated to the open-string partition function on the disk.

\chapter{Moduli spaces and their BV algebra}

In the present section we will begin by discussing
a few of the basic properties of the moduli spaces of Riemann
surfaces with boundaries and punctures. The punctures can be open
string punctures, if they are located at boundaries, and closed
string punctures, if they are located in the interior of the surface.
We then define the sewing of surfaces, and briefly discuss the
five inequivalent sewing operations that can be performed on surfaces
with open and closed string punctures.

 In order to define
an antibracket and a delta operator (of BV type) acting on moduli 
spaces we  introduce a $Z$ degree on moduli spaces.
In addition we introduce a strictly associative multiplication of
moduli spaces.  We show that the moduli spaces must satisfy
cyclicity conditions: up to well  defined sign factors, the
moduli spaces are invariant under the operation of cyclic permutation
of the labels of the punctures lying on any boundary component.
We define the antibracket and verify it satisfies all the desired
properties.   This is an extension of 
the definition of a BV algebra for the moduli spaces of surfaces without
boundaries [\senzwiebachtwo, \senzwiebachnew].  Some nodding familiarity
with the discussion of Ref.[\senzwiebachtwo] will be assumed.

\section{Moduli spaces for oriented open-closed string theory}

The general moduli space is that of (oriented) Riemann surfaces 
of genus $g\geq 0$, with $n\geq 0$ labelled 
interior punctures, representing closed
string insertions and $b\geq 0$ boundary components. At the boundary components
there are labelled
punctures representing open string insertions.
Let $m_i\geq 0$ denote the 
number of punctures at the $i$-th boundary component. 
The total number of  open string punctures 
is therefore $m = \sum_{i=1}^b m_i$. 
Except for some low dimensional cases the (real)
dimensionality of the moduli space $\M^{g,n}_{b,m}$ 
is given by 
$$ \hbox{dim}  \M^{g,n}_{b,m} = 6g-6+2n+3b+m \, . \eqn\dimmod$$

All punctures must be equipped with analytic local coordinates
(see Fig.~1)
As usual the closed string punctures are equipped with local
coordinates defined only up to phases. These are simply defined via
an analytic map of a unit disk into the surface, with the origin
of the disk going to the closed string puncture. There is no
natural way to fix the phase of the local coordinate at an interior
puncture. 

The local coordinates for the open string punctures are defined as
follows (Fig.~1). An open string coordinate
is an analytic map from the  upper half-disk 
$\{ |w|\leq 1, \hbox{Im}(w) \geq 0\}$ into a neighborhood
of the puncture, with the origin going to the puncture and the boundary
$\{ \hbox{Im} (w) = 0\}$ of the half-disk going into the boundary of
the Riemann surface. There is no phase ambiguity in this definition.
The real axis of the local coordinate coincides with the boundary, and
positive imaginary values are inside the surface.
Note that the orientation of
the Riemann surface, defined by the usual orientation on every
complex chart, induces an orientation on the boundary components,
an orientation that can be pictured as an arrow.
As we travel along the oriented boundary we always 
move along increasing real values for the local coordinates.

\Figure{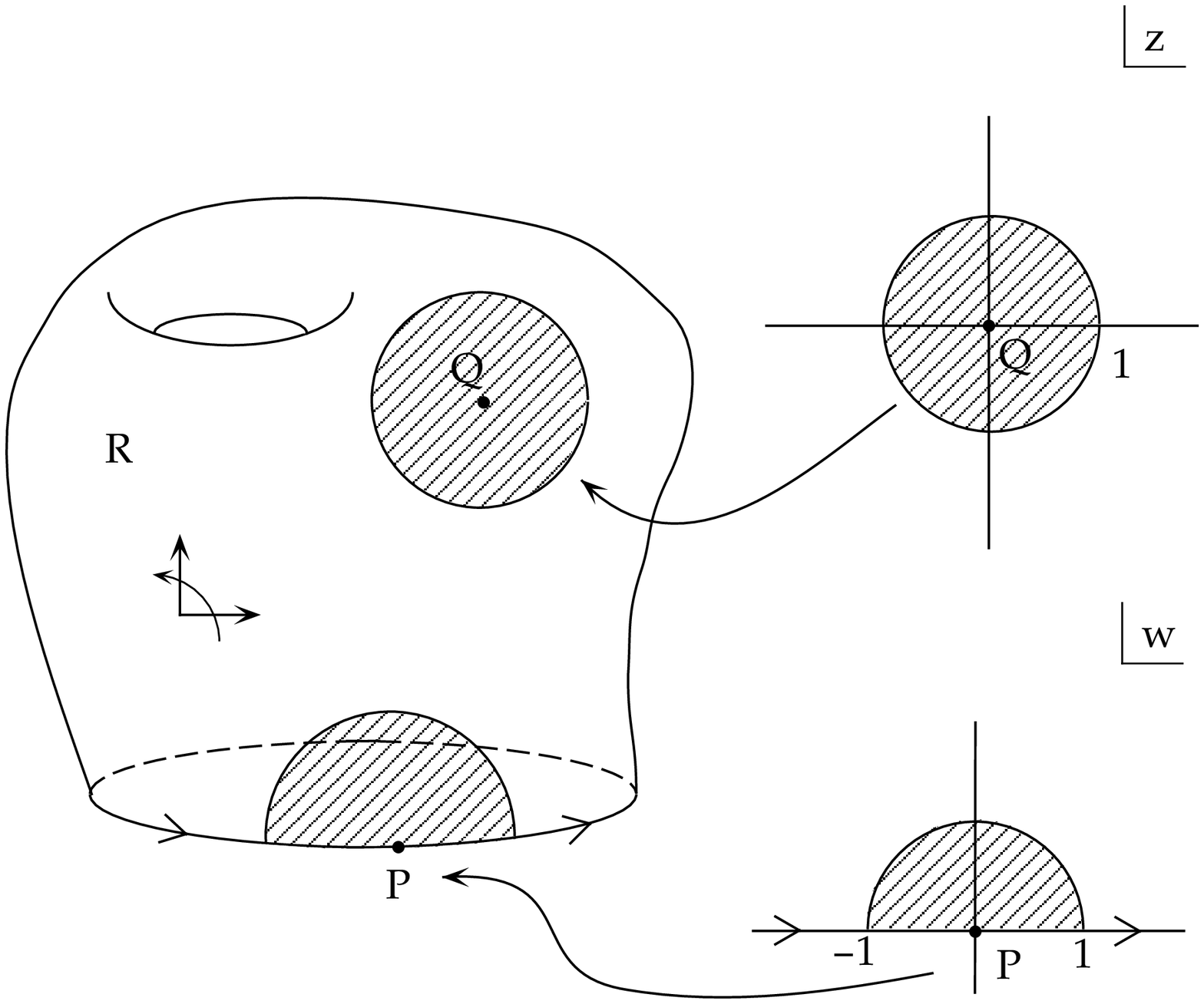 scaled 500}{Figure 1.~A Riemann surface $R$ with a closed string
puncture at
$Q$ and an open string puncture at $P$. The local 
coordinate $z$ at $Q$ is defined
by an analytic map from the unit disk to the surface. The local 
coordinate $w$ at $P$ is defined by an analytic map of the
upper-half-disk into the surface. The real line in the upper-half-disk
maps into the boundary of the surface $R$.}

It will be useful to have an ordered list of the lowest
dimensional moduli spaces of punctured surfaces with boundary
components. We do not assign a negative contribution
to the dimensionality arising from  conformal
Killing vectors. Our list begins with dimension zero moduli spaces.

\goodbreak
\noindent
\underbar{Dimension zero} 
\nobreak

--the sphere with zero${}^{*\,6}$, one${}^{*\,4}$, two${}^{*\,2}$, or
three closed string punctures;

--the disk with zero${}^{*\,3}$, one${}^{*\,2}$, two${}^{*\,1}$, or 
three open string punctures; 

--the disk with one${}^{*\,1}$ closed string puncture, and, 

--the disk with one open and one closed string puncture. 

\noindent
The surfaces followed by an asterisk as in $\{\,\,\}^{*\, n}$ have $n$
real conformal Killing vectors. 

\goodbreak
\noindent
\underbar{Dimension one}
\nobreak

--the disk with four open punctures,

--the disk with one closed and two open punctures, 

--the disk with two closed punctures, and,

--the annulus with zero${}^{*\,1}$ or one open string puncture.

\goodbreak
\noindent
\underbar{Dimension two}  
\nobreak

--the torus with zero${}^{*\,2}$ or one puncture,

--the four punctured sphere,

--the disk with five open punctures,

--the disk with one closed and three open punctures, 

--the disk with two closed and one open puncture, 

--the annulus with two open punctures (two cases), and,

--the annulus with one closed puncture.

\noindent
For the case of the annulus with two open punctures, the two punctures
may lie in the same boundary component, one puncture
may lie in each boundary component.

\section{Sewing of surfaces and moduli spaces}

An interaction vertex is represented by a blob (Fig.~2)
and will be accompanied by the data $(g,n,b,m)$.
Only if needed explicitly we will give the number
of open strings at each boundary component.
Wavy lines emerging from the  blob 
represent closed strings, each
heavy dot represents a boundary component, and the straight
lines emerging from them are open strings .

\Figure{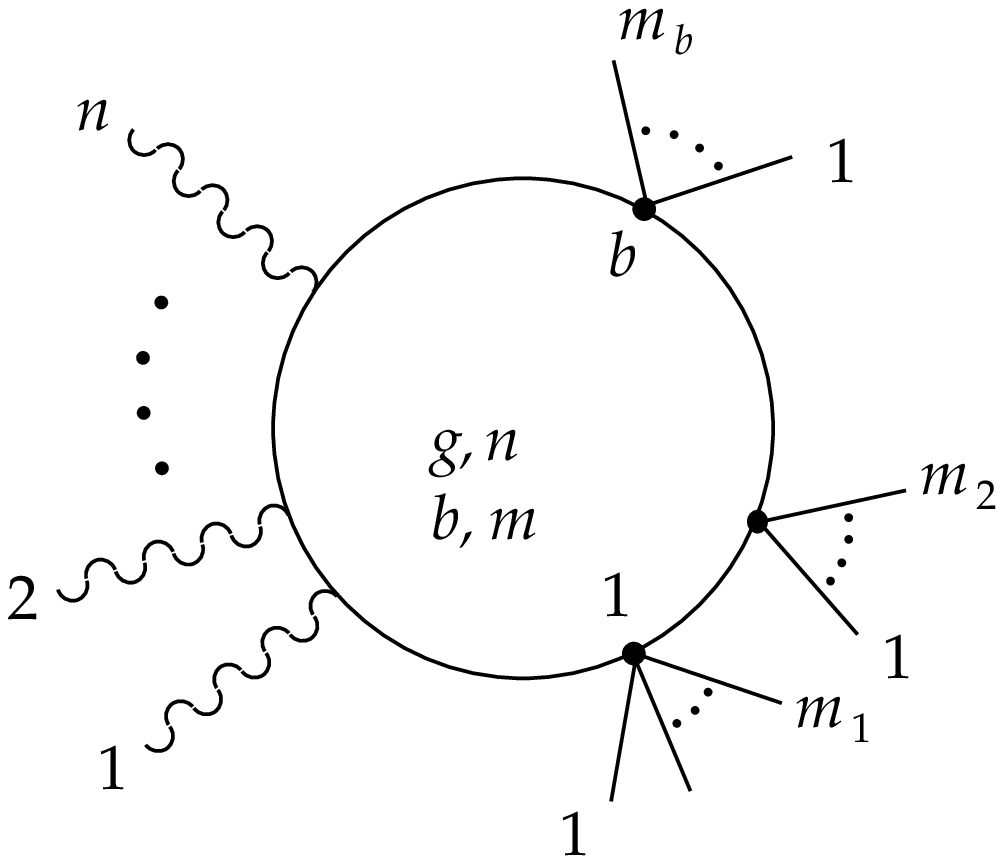 scaled 700}{Figure 2.~The representation of a string vertex arising from
the moduli space of surfaces of genus $g$, with $n$ closed string punctures,
$b$ boundary components, and a total of $m$ open string punctures. The
heavy dots represent boundary components. Straight lines are open string
punctures and wavy lines represent closed string punctures. 
The $i$-th boundary component has $m_i$ punctures, and $m= \sum_{i=1}^b m_i$.}

The  canonical sewing operation for open strings is described by an identification of
the type  
$z w =  t$, where $z$ and $w$ are two local coordinates defined for open string
punctures and $t$ is a constant.  Note that this constant should be real so that
the boundaries of the half-disks defined by the local coordinates are glued 
completely into each other. Moreover, given our definition of the half disks
as corresponding to the region of the disk with positive imaginary part, 
the point $z=i$ could only be glued with the point $w=i$, and this requires that the
constant be equal to minus one. Therefore, the canonical sewing
operation for open strings is of the form
 $$zw = -1   \, . $$
For closed strings, one usually takes the canonical sewing operation
to be defined as $zw=1$ where $z$ and $w$ are local coordinates around
closed string punctures.

Consider a single sewing operation. It may involve two surfaces or it may involve
a single surface. Moreover, the sewing operation may be the sewing of two open
string punctures or the sewing of two closed string punctures.  We want to verify
that whenever the sewing is of open string punctures the resulting surface belongs
to a moduli space whose  real dimensionality is greater by one unit than that
of the moduli space (s)  of  the original surface (s).  For closed string sewing, the
resulting surface belongs
to a moduli space whose real dimensionality is greater by two units than that
of the moduli space (s)  of  the original surface (s).  An obvious consequence of
the above statements is very familiar. Sewing of open string punctures with one real 
variable sewing parameter, and sewing of closed string punctures with two 
real sewing parameters, are both operations that  starting with moduli spaces
of proper dimensionality give moduli spaces of proper dimensionality.

Let us first consider sewing of open string punctures.  Here there are three 
possibilities:

\noindent
(i) The open string sewing joins two surfaces. In this case the genus and the
number of closed string punctures simply add. The total number of boundaries
is decreased by one, and the total number of open string punctures is decreased
by two. One readily verifies using \dimmod\ that 
$$ \hbox{dim}  \M^{g_1+ g_2\,  , \, n_1+n_2}_{b_1+b_2 -1, m_1 + m_2 -2}
=  \hbox{dim}  \M^{g_1, n_1}_{b_1, m_1} + 
 \hbox{dim}  \M^{g_2 , n_2}_{b_2 ,  m_2} + 1 \, .  \eqn\sewone$$

\noindent
(ii) The open string sewing joins two open string punctures lying on the same boundary
component of a single surface. In this case the genus and the
number of closed string punctures do not change. The total number of boundaries
is increased  by one, and the total number of open string punctures is decreased
by two. Again, one verifies that
$$ \hbox{dim}  \M^{g\,  , \, n}_{b+1, m -2}
=  \hbox{dim}  \M^{g, n}_{b, m} + 1 \, .   \eqn\sewtwo$$

\noindent
(iii) The open string sewing joins two open string punctures lying on different boundary
components of a single surface. This operation acturally increases the genus by
one unit and decreases the number of boundaries by one unit. One can understand
this as follows. Gluing completely two boundaries adds handle; partial gluing of two
boundaries is the same as adding a handle with a hole. Gluing of open string punctures
is indeed partial gluing of boundaries, thus explaining why the genus increases
by one, while the number of boundaries decreases by one.  A short  calculation 
confirms that
$$ \hbox{dim}  \M^{g+1\,  , \, n}_{b-1, m-2}
=  \hbox{dim}  \M^{g, n}_{b, m} + 1 \, .   \eqn\sewthree$$

For the case of sewing of closed string punctures there are only two configurations:

\noindent
(iv) Closed string sewing of two surfaces.  Here the number of  boundaries and
the number of open string punctures simply add.  The genus adds, and the total number
of closed string punctures is reduced by two.  One readily finds that  
$$ \hbox{dim}  \M^{g_1+ g_2\,  , \, n_1+n_2-2}_{b_1+b_2\, , \, m_1 + m_2}
=  \hbox{dim}  \M^{g_1, n_1}_{b_1, m_1} + 
 \hbox{dim}  \M^{g_2 , n_2}_{b_2 ,  m_2} + 2 \,.  \eqn\sewfour$$

\noindent
(v) Closed string sewing involving a single surface. Here the number of  boundaries and
the number of open string punctures remain the same.  The genus is 
increased by one unit, and the total number
of closed string punctures is reduced by two units.  One readily confirms  that  
$$ \hbox{dim}  \M^{g+1,  n-2}_{b\, , \, m}
=  \hbox{dim}  \M^{g, n}_{b, m} +   2 \,.  \eqn\sewfive$$

\section{Assigning a $Z$ degree to moduli spaces}

In order to build a Batalin-Vilkovisky algebra whose elements are moduli 
spaces of surfaces we must assign a  $Z_2$ grading to the moduli spaces.
In the closed string case this $Z_2$ grading was given by the dimensionality
of the moduli space in question.  The antibracket operation of two surfaces
amounted to twist
sewing (sewing with $zw= \exp (i\theta) , 0< \theta < 2\pi $) of two closed string
punctures one in each surface. The antibracket of two moduli spaces is the set
of surfaces obtained by taking the antibracket of every surface in the first moduli
space with every surface in the second moduli space. 
Since twist sewing has one real parameter, it gives a moduli space
whose dimension is one unit bigger than the sum of the dimensions of 
the moduli spaces that are to be sewn. This is exactly what
one wishes, since, by definitiion, the antibracket is an operation with odd degree.

This brings in a first puzzle.  It is quite clear from experience with open
string theory that the antibracket on the open
string sector can only amount to the sewing of two open string punctures 
lying on two different surfaces.  In this case the antibracket would not
change dimensions and would seem to correspond to an operation of even degree. 
It is clear that we need a new definition of the degree associated to a moduli space
of surfaces.

Let us first notice that  in the closed string case there was one important fact:
all moduli spaces of proper dimensionality had even degree. This was easily
achieved since the dimension of $\M^{g,n}$ is always even. We wish to have
the same property for  open-closed moduli spaces of proper dimensionality.  This
cannot be achieved by setting the degree equal to the dimension, since 
the dimension of $\M^{g,n}_{b.m}$ can be odd.  The way out is clear, for any
given moduli space $\A^{g,n}_{b,m}$ we define the  $Z$ degree $\e$
 to be given by 
$$\e (\A^{g,n}_{b,m} ) = 
\hbox{dim} \M^{g,n}_{b.m} - \hbox{dim} \A^{g,n}_{b.m} \, .\eqn\newdeg $$
Note that  for moduli spaces of closed surfaces this degree induces exactly
the same $Z_2$ degree we had before. Moreover, by 
definition, proper dimensionality
moduli spaces now have degree zero. If we only need a $Z_2$ degree,
many of the terms in \dimmod\ are irrelevant and we find
$$\e (\A ) = b + m + \hbox{dim} \A    \, . \eqn\diskgrade$$

More important, we can now verify 
easily that the antibracket will have degree equal to $+1$ both in the open and
in the closed string sector.  Consider two moduli spaces $\A_1$ and $\A_2$
whose antibracket we are computing in the open string sector. 
Let $\M_1$ and $\M_2$ denote
the canonical moduli spaces  associated to $\A_1$ and $\A_2$ respectively. Moreover,
let $\M_{12}$ denote the canonical moduli space associated to the surfaces that
appear in the open string antibracket $\{ \A_1 , \A_2\}_o$.  Since open 
string sewing adds no dimension, the dimension of $\{ \A_1 , \A_2\}_o$ is just
the sum of the dimensions of $\A_1$ and $\A_2$. It then follows that 
$$\eqalign{
 \e  ( \{ \A_1 , \A_2 \}_o ) &= \hbox{dim} \M_{12} - 
(\hbox{dim} \A_1  + \hbox{dim} \A_2), \cr
& = \hbox{dim} \M_1 + \hbox{dim} \M_2 + 1   -\hbox{dim} \A_1  - \hbox{dim} \A_2,  \cr
& = (\hbox{dim} \M_1 -\hbox{dim} \A_1  ) + (\hbox{dim} \M_2    -\hbox{dim} \A_2) + 1  ,  \cr
& =\e (\A_1  ) +\e (\A_2) + 1  ,  \cr  } \eqn\degworks$$
where use was made of  \sewone.  A rather analogous computation gives the same exact
result for  closed string antibracket $\{ \A_1 , \A_2\}_c$. 
This time, twist sewing implies
that the dimension of $\{ \A_1 , \A_2\}_c$ is equal 
to (dim $\A_1 + $ dim$\A_2 + 1$),
while $\hbox{dim} \M_{12} = \hbox{dim} \M_1 + \hbox{dim} \M_2  + 2$, 
by virtue of \sewfour .

We can also verify that the delta operation $\Delta$ is also an operation of degree
equal to $+1$.  Recall that in the closed string sector $\Delta$ twist sews two closed
string punctures lying on the same closed surface. In the open string sector the delta
operator becomes an operator $\Delta_o$ that will
simply sew two open string punctures lying on the same surface.\foot{The BV algebra
of surfaces can be described taking $\Delta$ to be the fundamental operation, as
done for closed Riemann surfaces in [\senzwiebachnew]. In here we will describe
explicitly the antibracket, and only describe $\Delta$ qualitatively.}  Denoting by 
$\M$ the canonical moduli space associated to $\A$ and by $\Delta_o \M$ the 
canonical moduli space associated to $\Delta_o\A$ we have
$$\eqalign{
 \e  (\Delta_o \A) &= \hbox{dim} \Delta_o\M - \hbox{dim} \Delta_o \A , \cr
& = \hbox{dim} \M +  1   -\hbox{dim} \A,  \cr
& = (\hbox{dim} \M -\hbox{dim} \A) + 1  ,  \cr
& =\e (\A_1  )  + 1  ,  \cr  } \eqn\degdworks$$
where use was made of either \sewtwo\ or \sewthree. An identical equation holds 
for the closed string sector of the delta operation. Finally, the boundary operator,
which decreases the dimensionality of a moduli space by one unit, also has 
degree $+1$
$$ \e (\partial \A ) = \e (\A ) + 1\, , \eqn\degpart$$ 
as is clear from the definition of degree.  All in all we have found a definition
of the $Z$ degree of moduli spaces such that the operations $\{ \cdot , \cdot \}$,
$\Delta$ and $\partial$ have all degree equal to plus one.  Moreover, moduli spaces
of proper dimensionality are all of degree zero, and thus correspond to even elements
of the algebra.

\section{The cyclic complex $\P$ of moduli spaces}

While we know from the previous subsection what is roughly the definition of
the antibracket,  we must define carefully its action in order 
to verify that it satisfies
the correct exchange property
$$\{ \A_1 , \A_2 \} = - (-)^{(\A_1 + 1 ) (\A_2 + 1)} \, \{ \A_2 , \A_1\} \, , 
\eqn\antexch$$
as well as the Jacobi identity
$$ (-)^{(\A_1 + 1 ) (\A_3 + 1)} \, \{ \A_1 ,  \{  \A_2 ,  \A_3  \} \,  \}  + \, 
\hbox{cyclic} = 0 \, . 
\eqn\jacobii$$

Our strategy will be to consider first the case when the moduli spaces contain
surfaces having a single boundary component and we will look at the 
antibracket in the open string sector.  Both $\A_1$ and $\A_2$ are oriented,
with orientations defined by  a set of basis vectors $[\A_1]$ containing dim $(\A_1)$ 
vectors, and a set of basis vectors $[\A_2]$ containing dim $(\A_2)$ 
vectors, respectively. 
We will define the orientation of the antibracket $\{ \A_1 , \A_2\}_o$ 
by the ordered set of
vectors $[ [\A_1 ] ,[\A_2]]$. 
We face a puzzle in trying to reproduce \antexch.  It seems faily clear that
the antibracket $\{ \A_1 , \A_2\}_o$ and the antibracket $\{ \A_2 , \A_1\}_o$
could only be related by a sign factor involving the product of the dimensions of the
respective moduli spaces. If this is the only sign factor, one cannot reproduce the
sign factor involving the product of the degrees, since the degrees involve the
number of open string punctures, as shown in \diskgrade.

The solution to this  complication is interesting. We have to work in a complex where
the moduli spaces have suitable properties under a change
of labels associated to the  the open string punctures. 
 Indeed, in the closed string sector, moduli spaces are assumed to be
invariant under any permutation of labels associated to closed string punctures.
For moduli spaces of bordered surfaces we introduce a {\it cyclic complex}  $\P$, where
the word cyclic refers to a property under the cyclic permitation of open string
punctures on a boundary component.  Consider a surface
having a single boundary component and $m$ open string punctures located at
this boundary.  The punctures will be labelled in cyclic order, namely, as we
go around the oriented boundary the labels of the punctures we encounter
always  increase by one unit (mod $m$).  
Let $\C$ denote the operator that acting on a surface $\Sigma$ 
gives the surface $C\Sigma$ which differs from $\Sigma$ by a cyclic 
permutation of the labels 
of the punctures. After the action of $C$ the puncture that used to have the label $1$
has the label $2$, the puncture that had label $2$ now is labeled $3$ and so on.
If we denote by $l_P(\Sigma)$ the label of the point $P$ in $\Sigma$, this means
that
$$l_P(C\Sigma) = l_P(\Sigma) + 1  \quad [\hbox{mod} \, m]  \, . \eqn\modm$$
Acting on a moduli space of surfaces the operator $C$ will do a cyclic permutation
of the punctures in every surface of the moduli space.  It is clear that by definition
acting on a space with $m$ punctures $C^m = 1$.  A moduli space $\A$ of surfaces
having  a single boundary component  and $m$ punctures  is
said to belong to the cyclic complex $\P$ if 
$$C \A  = (-)^{m-1} \A \,,  \quad \forall \A \in \P\, . \eqn\belong$$
Namely, under
a cyclic permutation of the labels, 
the moduli space goes to itself except for the sign factor 
$(-)^{m-1}$.  This nontrivial sign factor will play a role in giving the correct
exchange property for the antibracket. Note that \belong\ is consistent
with $C^m \A = \A$.  The sign factor for a cyclic step is 
the sign factor that would arise if we do the cyclic step by successive exchanges
of pairs of labels, and we assume that each exchange carries a factor of minus one.
This is in accord with the idea that the open string field is naturally odd in some
formulations of open string theory. In the present formulation, where the open
string field is even, the odd property is carried by the complex of moduli space.
Indeed, the symplectic form, associated to a two-punctured disk, is naturally
odd in the present formulation.  Given a moduli space $\A$ having no particular
cyclicity property  we will denote by 
$$ (\A)_{cyc} \equiv \sum_{i=1}^{m-1}   (-)^{i(m-1)} \,  C^i  \A \eqn\cycliccom$$
the  cyclic moduli space that arises by explicit addition of $m-1$ 
cyclic copies of $\A$
weighted with the appropriate sign factors. It is simple to verify that $(\A)_{cyc}$
satisfies \belong.  
 
\section{An associative multiplication}  

It will be helpful to introduce 
a multiplication of surfaces (each having a single boundary component, for 
simplicity). Given a surface $\Sigma_1$ and a surface $\Sigma_2$ we
define the product $\Sigma_1 \circ \Sigma_2$ as the surface obtained
by gluing the last puncture of $\Sigma_1$ to the first puncture of $\Sigma_2$.
The remaining punctures are relabeled cyclically by keeping the original
labels in the remaining punctures in $\Sigma_1$ and extending this
to the glued surface.  It is important to notice that the first puncture of 
$\Sigma_1 \circ \Sigma_2$ is the first puncture of $\Sigma_1$, and the
last puncture of $\Sigma_1 \circ \Sigma_2$ is the last puncture of $\Sigma_2$.
This multiplication is not commutative, but it is clearly strictly associative.
This multiplication is extended in the obvious way to a multiplication of 
moduli spaces, preserving associativity. The orientation of the moduli space
$\A\circ \B$ is defined by the set of vectors  $[ [\A] [\B]]$ where $[\A]$ and
$[\B]$ are the vectors defining the orientations of $\A$ and $\B$ respectively.
Note, however, that if  two
moduli spaces are cyclic, the product will not be.  This is clear, because
in all of the resulting surfaces the last puncture always  lies on 
the part of the boundary
component that originated from surfaces in the second moduli space, and it is
always one step away from the place where the sewing operation takes place. This
is clearly not a cyclic moduli space.  I have not found a way to define
an associative multiplication in the cyclic complex.

\Figure{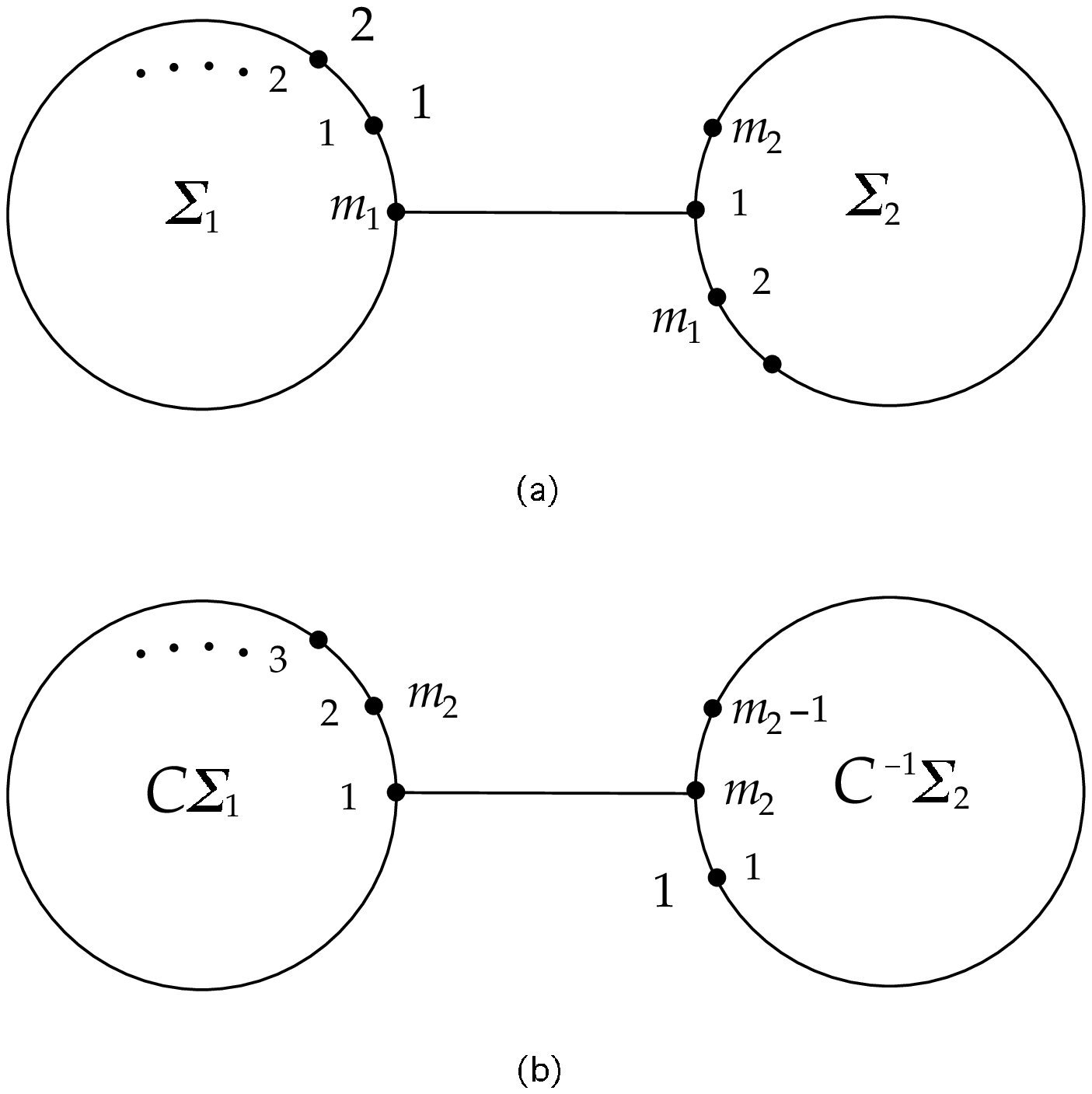 scaled 650}{Figure 3.~ (a) The product  $\Sigma_1 \circ\Sigma_2$
joins the last puncture of  $\Sigma_1$ (label $m_1$)  to the first puncture
of $\Sigma_2$.  The labels  inside the blobs are the original labels of the
punctures, the labels outside the blobs are the labels after gluing. 
(b) The product  $C^{-1}\Sigma_2 \circ C\Sigma_1$.
Note that  the final labels of the punctures in cases (a) and (b) 
can be made to agree by doing
$m_2-1$ cyclic permutations of the punctures in (b). }

While the multiplication is not commutative, by suitable cyclic permutations
of the punctures we can obtain an equality relating the two different ways of
multiplying two surfaces. We claim that
$$\Sigma_1 \circ \Sigma_2 =  
(C^{-1})^{m_2 -1}  \bigl(  C^{-1}\Sigma_2 \circ C\Sigma_1\bigr) \eqn\commu$$
This property is best explained making use of the Fig.3. Note that in the left
hand side we use the last puncture of $\Sigma_1$ while the operation on the
right hand side would use the first puncture of $C\Sigma_1$. This is as it
should be since these punctures are really the same concrete``physical'' puncture
(see \modm).
Similarly, the first puncture of $\Sigma_2$ is the same puncture as the last puncture
of $C^{-1}\Sigma_2$. The overall  cyclic factor $(C^{-1})^{m_2-1}$ in the right hand 
side is necessary because the labels of the resulting surfaces should also agree.
In the left hand side the first puncture of the glued surface is the first puncture of
$\Sigma_1$, while in the right hand side the same puncture carries the label
$m_2$.  It follows that $m_2-1$ anti-cyclic permutations of the right hand side
are necessary for these
labels to coincide. This concludes the verification of  equation \commu. 
When we deal with moduli spaces, there is an extra sign factor in the above relation.
We have that 
$$\A_1 \circ \A_2 =  (-)^{d_1 d_2} 
(C^{-1})^{m_2 -1}  \bigl(  C^{-1}\A_2 \circ C\A_1\bigr)\,,  \eqn\commum$$
where $d_1$ and $d_2$ denote the dimensions of the moduli spaces 
$\A_1$ and $\A_2$ respectively.  This  sign factor arises because of 
the way we defined  the orientation for the product of two moduli spaces. 
If the moduli spaces involved are cyclic, that is,   $\A_1, \A_2 \in \P$,
the above exchange property simplifies considerably
$$\A_1 \circ \A_2 =  (-)^{d_1 d_2 + m_1 + m_2 } 
(C^{-1})^{m_2 -1}  \bigl(  \A_2 \circ \A_1\bigr) \, \eqn\commuxm$$
where the extra factors arise from the action of $C$ and $C^{-1}$ on the cyclic
moduli spaces. In order to find an equation that does not involve
explicitly the operator $C$  we can now  form the
cyclic moduli spaces as in equation \cycliccom\
to find
$$(\A_1 \circ \A_2)_{cyc} =  (-)^{d_1 d_2 + m_1 + m_2 + (m_2-1) (m_1 + m_2 -3) } 
 \bigl(  \A_2 \circ \A_1\bigr)_{cyc}\,, \eqn\commuy$$
where the new sign factor arises because the moduli space 
$(  \A_2 \circ \A_1)_{cyc}$ is cyclic and its surfaces have $(m_1+m_2-2)$
punctures. The sign factor can be simplified to read
$$(\A_1 \circ \A_2)_{cyc} =  -(-)^{d_1 d_2 + m_1m_2}  
\bigl(  \A_2 \circ \A_1\bigr)_{cyc} \,.\eqn\commuyy$$
Note that this exchange property involves not only the dimensionalities
of the moduli spaces but also the numbers of punctures. This is the kind
of result that we
needed for the antibracket, since the degree of a moduli space involves
both the dimension of the space and the number of open string punctures.

\section{Defining the antibracket}

The exchange property of the antibracket, given in \antexch\ requires
a sign factor that differs slightly from that in \commuyy. The desired
sign factor arises if we define the antibracket 
(in the open string sector) as 
$$\{\A_1 , \A_2 \}_o \equiv  (-)^{m_1d_2} \, (\A_1 \circ \A_2 )_{cyc}\, .\eqn\prant$$ 
It is now a simple computation using \commuyy\ to verify that
$$\{ \A_1, \A_2\}_o =  -(-)^{(d_1 +m_1)(d_2 + m_2)}  \{ \A_2 , \A_1\}_o \,.\eqn\cuyy$$
Using \newdeg\ and \dimmod\ we see that the $Z_2$ degree of
a moduli space with one boundary component is $\e (\A) = m+  d_\A + 1$.
It then follows that \cuyy\ is the correct exchange property 
for the antibracket. 

The antibracket must satisfy a Jacobi identity of the form indicated in \jacobii.
Consider the first term in this identity (always in the open string sector
of the antibracket). Using \prant\ we find that 
$$ (-)^{(\A_1 + 1 ) (\A_3 + 1)} \, \{ \A_1 ,  \{  \A_2 ,  \A_3  \} \,  \} 
 = (-)^{s_{13}}\bigl(\A_1 \circ ( \A_2 \circ  \A_3  )_{cyc} \bigr)_{cyc}
 \, ,  \eqn\jac$$
where the sign factor $s_{13}$ is given by
$$s_{13} = d_1 d_3 + m_1 m_3 + (d_1 m_3 + d_2 m_1 + d_3 m_2 ) \,. $$
On the other hand, the last term of the Jacobi identity, for example, would read
$$ (-)^{(\A_3 + 1 ) (\A_2 + 1)} \, \{ \A_3 ,  \{  \A_1 ,  \A_2  \} \,  \} 
 = (-)^{s_{32}}\bigl(\A_3 \circ ( \A_1 \circ  \A_2  )_{cyc} \bigr)_{cyc}
 \, ,  \eqn\jacc$$
where the sign factor $s_{13}$ is given by
$$s_{32} = d_3 d_2 + m_3 m_2 + (d_1 m_3 + d_2 m_1 + d_3 m_2 )\, . $$
We wish to verify that surfaces with the same sewing structure in \jac\ and \jacc\
appear with opposite signs. Given three surfaces $\Sigma_1, \Sigma_2,$ and $\Sigma_3$
belonging to $\A_1,\A_2, $ and $\A_3$ respectively, we can use the 
associativity of the $\circ$-product, and equation \commu\  to write
$$\eqalign{
\Sigma_1 \circ (\Sigma_2\circ\Sigma_3) &= (\Sigma_1\circ\Sigma_2)\circ\Sigma_3\,,\cr
&= (C^{-1})^{m_3-1} \bigl( C^{-1} \Sigma_3 \circ  
C ( \Sigma_1 \circ \Sigma_2)\bigr) \, .\cr} \eqn\getjac$$
This equation implies that configurations in the right hand side of \jacc\
having the same structure as those appearing in the right hand side of \jac\
have an additional sign factor
$$(m_3\hskip-2pt -\hskip-2pt 1) (m_1\hskip-2pt +\hskip-2pt m_2\hskip-2pt +
\hskip-2pt m_3\hskip-2pt -\hskip-2pt 1) + (m_3\hskip-2pt -\hskip-2pt 1) + 
(m_1\hskip-2pt +\hskip-2pt m_2\hskip-2pt +\hskip-2pt 1 ) = m_3 m_1\hskip-2pt +
\hskip-2pt
m_3 m_2 \hskip-2pt + \hskip-2pt 1 \quad (\hbox{mod}\, 2)\, . $$
We then confirm that
$$s_{32} + d_3(d_1 + d_2) + m_3 m_1\hskip-2pt +
\hskip-2pt
m_3 m_2 \hskip-2pt + \hskip-2pt 1 = s_{13} + 1\,, $$
thus completing the verification that identical surfaces appearing in
the various terms of the Jacobi identity cancel out as desired.

Another important property of the antibracket must be its behavior under
the action of the boundary operator $\partial$.  For our multiplication we have
$$\partial (\A_1 \circ \A_2)_{cyc} = 
 (\partial \A_1 \circ \A_2)_{cyc}  + (-)^{d_1}  (\A_1 \circ \partial\A_2)_{cyc} \,, \eqn\parta$$
and making use of \prant\  the above equation reduces to
$$\partial \{ \A_1 ,  \A_2\}_o = 
\{ \partial \A_1 ,  \A_2\}_o  + (-)^{\A_1 + 1}  \{ \A_1 ,  \partial\A_2\}_o\,, \eqn\partaa$$
which is the expected result.

\section{Multiple boundary components}

We now sketch briefly how the above results should be extended when 
the complex $\P$ contains moduli spaces $\A$  whose surfaces
have more than one boundary component. 
In this case the moduli space $\A$  will have to satisfy further exchange
properties. Let $b$ denote  the number of boundary components  in  each of  the
surfaces contained in $\A$.  The boundaries must be labelled, and 
$\Gamma_k$, with $k$ running from one up to $b$, will denote the 
$k$-th boundary component.
We also let  $m_k$ denote the number of open string punctures on the boundary 
component $\Gamma_k$. The open string  punctures must be labelled cyclically
on each boundary component.
We denote by $C_k$ the operator that generates
a cyclic permutation of the punctures in $\Gamma_k$.  The moduli space $\A$ must
satisfy the cyclicity condition \belong\ for every boundary component:
$C_k \A = (-)^{m_k-1} \A, \, k=1,\cdots , b$.  Let now $E_{ij}$, for $i\not= j$
denote the operator
that exchanges the labels of the boundary components $\Gamma_i$ and $\Gamma_j$.
A moduli space $\A$ will belong to $\P$, if in addition to the previous constraints
it satisfies 
$$ E_{ij} \, \A = (-)^{m_i + 1)(m_j+1)}\,  \A  \, \, .\eqn\bexch$$
If a moduli space does not satisfy the above equation, one can always
construct a moduli space that does by adding together, with appropriate
sign factors,  the $b \, !$ inequivalent labellings of the boundaries. 

The multiplication ``$\circ$'' of surfaces is modified in a simple way preserving
associativity.  This time
$\Sigma_1 \circ \Sigma_2$
denotes the sewing of the last puncture of the last boundary component of
$\Sigma_1$ (the component $\Gamma_{b_1}^{(1)} $, where the superscript refers
to the surface)   to the first puncture of the first boundary component of 
$\Sigma_2$.(the component $\Gamma_{1}^{(2)}$).  The punctures in the
now common boundary component are labelled as we did before. This boundary
component is now the component $\Gamma_{b_1}^{(12)}$ of the sewn surface.
For $k< b_1$ we define $\Gamma_k^{(12)} = \Gamma_k^{(1)}$ and for  
$b_1 < k < b_1+ b_2 -1$
we let $\Gamma_k^{(12)} = \Gamma_{1+k-b_1}^{(2)}$. In words, the 
labels of the untouched
boundaries in $\Sigma_1$ are used for the sewn surface, and the boundary components
arising from the remaining boundaries on $\Sigma_2$ are labelled in ascending order.

Once more we need to quantify the failure of commutativity. Let $R$ denote the operator
that reverses the labels of all the boundaries namely $R: \Gamma_k \to \Gamma_{b+1-k}$.
The first boundary becomes the last, the last becomes the first, and so on.
We now claim that \commu\ gets generalized into the following expression 
$$\Sigma_1 \circ \Sigma_2 =  R\, ( C^{-1}_{b_2})^{m_1^{(2)}-1} 
\bigl(  R C_{1}^{-1} \Sigma_2
\circ  R C_{b_1}\Sigma_1) \, .\eqn\getcase$$
Note that the $C$ operators act on particular boundary components,
as indicated by the subscripts. The operators $R$ are necessary inside the
parenthesis in order to reverse last and first punctures. The operator
$R$ outside the parenthesis achieves an ordering of the punctures in the sewn surface
that coincides with that of the left hand side.

Given two moduli spaces $\A_1 , \A_2 \in \P$ we define
$$(\A_1 \circ\A_2)_* \in \P \, \eqn\spacee$$
as the space obtained by doing the product, adding all the cyclic 
permutations on the boundary component where the sewing took place,
and then adding with appropriate signs, copies of the space with
boundary labels exchanged such that the total spaces satisfies \bexch. 
Since both $\A_1$ and $\A_2$ satisfy \bexch\ it is sufficient to add over
all possible labels for the boundary that was sewn, and once this label is
chosen, we sum over splittings of the remaining labels into two groups, one
for the left over boundary components of $\A_1$ and the other for the
left over boundary components of $\A_2$.
We now claim that \getcase\ implies that
$$(\A_1 \circ\A_2)_*  = (-)^{s_{12}}\, (\A_1 \circ\A_2)_* \,, \eqn\rer$$
where the sign factor is given by
$$s_{12} = d_1d_2 + [1  \,+ (1+ b_1 + m^{(1)}) 
(1+ b_2 +  m^{(2)})] \,, \eqn\signmess$$
where the first term in the sign factor originates from the orientation,
the second term originates from the action of the $C$ and $R$ operators
indicated in \getcase. Here $m^{(1)}$ and $m^{(2)}$ denote the
total number of punctures in the surfaces belonging to $\A_1$ and $\A_2$
respectively.  It is now possible to define the antibracket
$$\{\A_1 , \A_2\}_o \equiv  (-)^{(1+b_1 + m^{(1)})}(\A_1 \circ\A_2)_*\,,\eqn\antg$$
and we readily verify that
$$\{\A_1 , \A_2\} _o = -(-)^{ (d_1+ 1+ b_1 + m^{(1)}) 
(d_2 + 1+ b_2 +  m^{(2)})} \{\A_2 , \A_1\}_o\, . \eqn\gfd$$
This is exactly the desired result, on account of \diskgrade\ and \antexch.

\chapter{Conditions on the string vertex $\V$}

In this section we set up and discuss the recursion relations for the string
vertices. They take the form of the BV equation for a string vertex
$\V$ representing the formal sum of all string vertices. We use this equation
to derive the $\hbar$ and $\kappa$ dependence of interactions.

We 
single out from $\V$ special  subfamilies of string vertices having
simple recursion relations. Those are: 

\noindent
(i)  the vertices corresponding
to surfaces without boundaries,

\noindent
 (ii)  vertices coupling all numbers
of open strings on a disk,

\noindent
 (iii) vertices coupling all numbers of
open  strings to $n\leq N$  closed
strings via a disk,
 together with  the vertices coupling $n\leq N+1$ closed strings on
a sphere, and

\noindent
 (iv) vertices coupling all numbers of open and closed
strings via a disk, together with vertices coupling all numbers of 
closed strings via a sphere.

\section{Master equation for String Vertices}

As in the closed string case we now consider a string vertex $\V$
defined as a formal sum of string vertices associated to the various
moduli spaces
$$\V \equiv \sum_{g,n,b,m}\,\,\hbar^p  \, \kappa^q
\sum_{m_1,\cdots m_b} \V^{g,n}_{b,m}\,\,. 
\eqn\sumv$$
All the moduli spaces $\V^{g,n}_{b,m}$ listed above are spaces of degree zero
(see \newdeg), and therefore they have the same dimension as $\M^{g,n}_{b,m}$.
Here the power $p$ of $\hbar$ and the power $q$ of the string
coupling associated to each string vertex are to
be determined as functions of $g,n,b$, and $m$. 
As the above expression indicates, having chosen $g,n,b$ and $m$,
one must still sum over the 
inequivalent ways of splitting the $m$ (labelled) open string punctures over
the $b$ boundary components. 

There are various moduli spaces that will not be included in the
above sum

-- the spheres ($g=b=0$) with $n\leq 2$,

-- the torus ($g=1$, $b=0$) with $n=0$, and,

-- the disks ($b=1$, $g=0$) with $m\leq 2$.

The idea now is to examine the recursion relations between
the moduli spaces that follow from
$$\partial \V + \half \{ \V \,, \V \} + \hbar\Delta \V = 0 \,,\eqn\recmot$$
where we impose this condition having in mind that it will lead
to the string action satisfying the BV master equation. We introduced a power of
$\hbar$ in \sumv\ in order to account for that factor of $\hbar$ appearing in front
of the delta operator.  The power of the string coupling was introduced  to see 
what are the conditions that \recmot\ implies on the string coupling dependence
of the interactions. 

It is best to discuss the various terms in the above equation
by referring to  figure 4. In the left hand side we have the boundary of
the general vertex defined by $(g,n,b,m)$. In the right hand side we have
five type of configurations. These are, in fact, the same 
configurations described in section 2.2. The first three configurations correspond
to sewing of open string punctures, and the last two configurations
correspond to sewing of closed string punctures, as familiar in the
closed string case. The configurations involving two vertices arise
from the antibracket, and the configurations involving a single vertex
arise from the delta operation. Let us examine each in turn.

\Figure{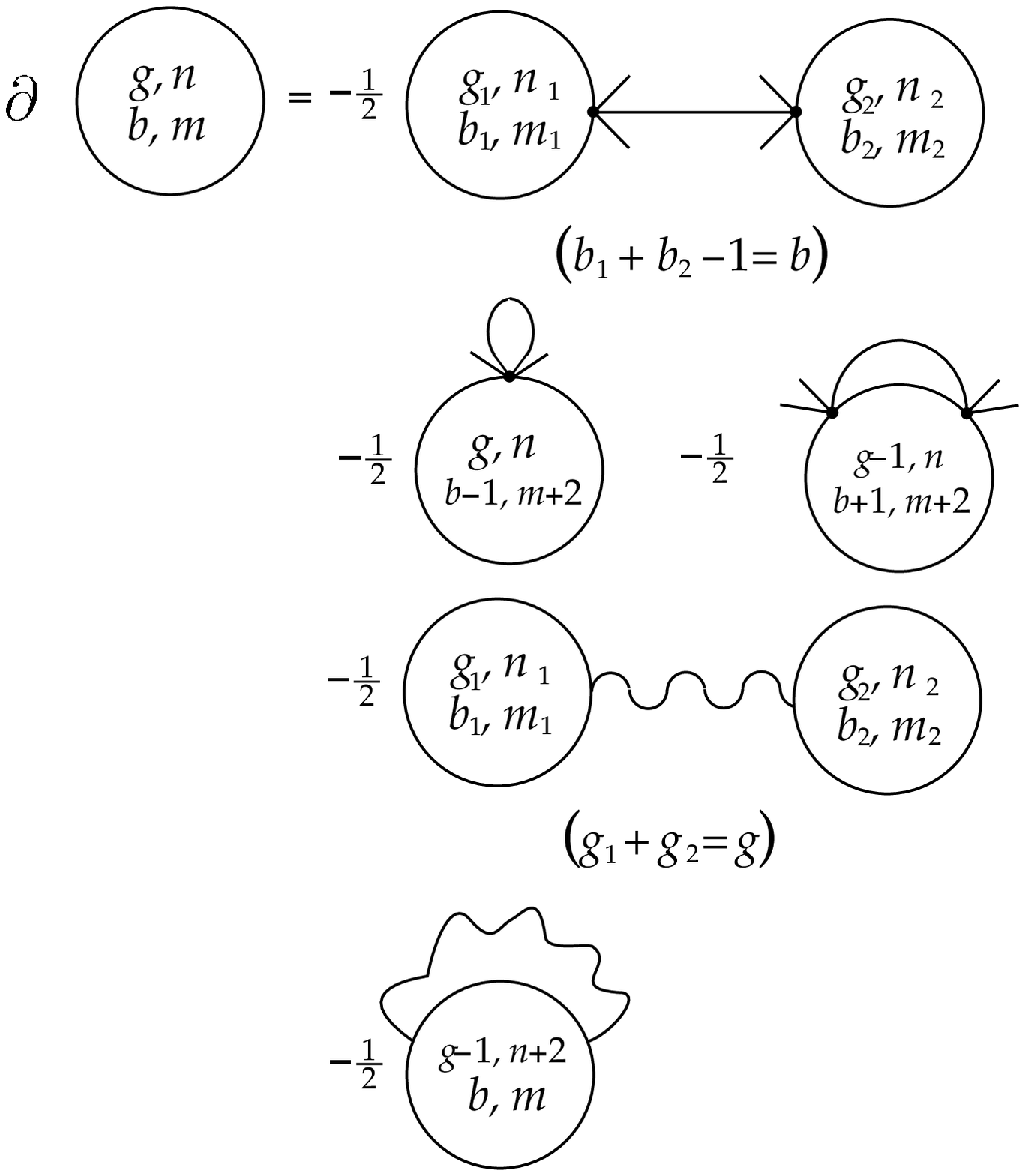 scaled 800}{Figure 4.~The master equation for open closed
string vertices. In the left hand side we have the boundary of 
the region of moduli space covered by a generic vertex. In the
right hand side we have the five inequivalent configurations featuring
a single collapsed propagator. The first three configurations arise
from sewing of open string punctures, and the last two configurations
arise from sewing of closed string punctures.}

(i) The first configuration, arising from the antibracket in the open 
string sector,  shows the gluing of 
open strings lying on boundary component of two 
different vertices. In this configuration the two boundary
components merge to become one, and therefore the total number 
$b$ of boundaries
in the glued surface is $b= b_1 + b_2 -1$. In addition,
$m= m_1+m_2-2$. 

(ii) The second configuration arises from the delta operation. Here the
two open strings to be glued are in the same boundary component of the
string vertex.
This operation increases the number of boundary components by one since the
gluing splits the  boundary in question into two disconnected components. The
number of open string punctures is decreased by two units.
The relevant blob indicated in the figure must therefore be of
type $(g,b-1,n,m+2)$.

(iii) The third configuration also arises from the delta operation. Here  the
two open strings to be glued are in different boundary components of the
string vertex. Just as in the case of configuration (i) 
the two boundaries involved become a single one. As explained in section 2.2, 
the genus increases by one. Thus the relevant blob for this
case must be of type $(g-1,n,b+1,m+2)$.

(iv) The fourth configuration arises from the antibracket in the closed string
sector. As is familiar by now this configuration requires that
the genera add to $g=g_1+ g_2$. The closed string punctures must satisfy
$n= n_1 + n_2 -2$, the  boundary components must
add to $b=b_1+b_2$ and the open string punctures must add to $m=m_1+m_2$.

(v) The fifth and final configuration arises from delta in the closed
string sector. Here the genus increases by one unit, the number of closed
string punctures is reduced by two units and both the number of boundaries
and the number of open string punctures are unchanged. 

It is straightforward
to verify that all the terms in the geometrical equation  are of the same dimensionality.
This was guaranteed by our construction. Since the antibracket, delta, and the
boundary operator have all degree one, all moduli spaces appearing in \recmot\
are guaranteed to have dimension one less than that of the corresponding moduli
space $\M$ . It follows that moduli spaces in \recmot\  of the same type (same $g,n,b,m$)
must have the same dimension.

\section{Calculation of $p$ and $q$}

We will now see that the values of $p$ and $q$ appearing in \sumv\
are severely constrained by the master equation \recmot. If we demand
that the three open string vertex appear in the action at zero-th order
of $\hbar$ and multiplied by a single power of the 
coupling constant $\kappa$ this will fix completely the values of $p$
and $q$ for all interactions. In order to see this clearly it is useful
to consider first the sewing properties of the Euler number $\chi$
associated to a surface of type $(g,n,b,m)$. We have
$$\chi\, (\Sigma^{g,n}_{b,m}\,) = 2 - 2g -n -b - \half m \,.\eqn\eulerno$$
This formula deserves some comment. The closed string punctures here
are treated simply as boundary component (note that $n$  appears 
in the same way as $b$). Each open string puncture is treated as 
half of a closed puncture, this is reasonable on account that upon doubling
the open string puncture would become a closed string puncture.

There are two important properties to $\chi$.
It is additive under sewing of two
surfaces
$$\chi(\Sigma_1\cup \Sigma_2) = \chi (\Sigma_1) + \chi (\Sigma_2)\,. 
\eqn\sewone$$
This is readily verified by checking the only two possible 
cases; sewing of closed string punctures,  and sewing of open string punctures.
It is moreover invariant under sewing of two punctures on the same 
surface
$$\chi (\cup \Sigma  ) = \chi (\Sigma)  \,, \eqn\sewto$$
and in this case there are the three cases to consider; sewing of two
closed string punctures, and sewing of two open string punctures, on the
same or on different boundary components. Note that the configurations
relevant to the two equations above all appeared in the recursion 
relations (Fig.4 ) and were discussed earlier. It is not hard to show
that conditions \sewone\ and \sewto\ fix the function $\chi$ to take
the form quoted in \eulerno\ up to an overall multiplicative constant.

Given a moduli space $\A^{g,n}_{b,m}$
of surfaces of type $(g,n,b,m)$ we will define $\ov \chi (\A) $ to be
given by the Euler number $\chi$ of the surfaces composing the moduli space
($\ov\chi$ is {\it not} the Euler number of $\A$).
We then have that the last two equations imply that
$$\eqalign{\ov\chi\,(\{ A_1,\A_2\}) &= \ov\chi\, 
(\A_1) + \ov\chi\, (\A_1)\,,\cr
\ov \chi \,(\Delta \A ) &= \ov \chi \,(\A)\,. \cr} \eqn\ontheab$$

We can now go back to discuss the problem of finding $p$ and $q$.
Let us begin with $q$, the power of the string coupling.  Assume we
study the subsector of the master equation having to do with
moduli spaces of some fixed type $(g,n,b,m)$. Such moduli space
appears in $\partial \V$, and appears built through an antibracket
in $\{\V,\V\}$ and by $\Delta$ action in $\Delta\V$. Since the 
string coupling appears nowhere in the master equation, all such
terms must have the same factor $\kappa^q$, and therefore,
 it is necessary that the assignment of $q$ to a moduli space  be
additive with respect to the antibracket, and invariant under
$\Delta$, namely 
$$\eqalign{q\,(\{ \A_1,\A_2\}) &= q\, 
(\A_1) + q\, (\A_1)\,,\cr
q\,(\Delta \A ) &= q\,(\A)\,. \cr} \eqn\oqth$$
These are exactly the same as \ontheab\ and therefore
the most general solution is $q= C \ov\chi$, where $C$ is a constant
to be determined. We require that $q$ be equal to one for the disk with
three open string punctures. For this surface $\ov\chi = -1/2$, and
therefore we find
$$q= -2 \,\ov\chi = 4g + 2n + 2b+ m -4\,. \eqn\detq$$

Let us now discuss the power $p$ of $\hbar$.
The master equation has an explicit factor of $\hbar$ in the
$\Delta$ term, therefore this time we must require that 
$$\eqalign{p\,(\{ A_1,\A_2\}) &= p\, 
(\A_1) + p\, (\A_1)\,,\cr
p \,(\Delta \A ) &= p \,(\A)\, + \, 1\,. \cr} \eqn\opth$$
It is clear that if we find any solution $p$, then we can
construct many solutions as $p + C\, \ov\chi$. In fact, one
can show explicitly that this is the most general solution. Therefore,
it is enough to find a solution of \opth. Define
$$\ov p (\A ) = 1- \half (n+ m) \,, \eqn\newfct$$
for a moduli space of type $(g,n,b,m)$. It is clear that  $\ov p$ 
solves \opth. It then follows that the most general solution is of the
type $p = \ov p + C\chi$. We require that $p=0$ for the disk with
three open string punctures. This fixes
$$p = -\ov\chi + \ov p = 2g + \half n + b -1 \,. \eqn\fixp$$
We have therefore found that the string vertex in \sumv\ is of the form
$$\V \equiv \sum_{g,n,b,m}\,\,\hbar{}^{-\ov\chi + \ov p}  \, 
\kappa{}^{-2\ov\chi}
\sum_{m_1,\cdots m_b} \V^{\, g,n}_{\, b,m}\,\,. 
\eqn\sumdv$$
Since the string action, except for the kinetic terms will just
be $f(\V)$, where $f$ is a map to the functions on the total state space of the
conformal theory, the above equation gives the order of $\hbar$ for
all the interaction terms in the string action.
The kinetic terms in the string action, 
both for the open and
closed string sectors appear at $\hbar^0$. Even though 
they should not be thought
as vertices, this is the value
of $p$ that follows from equation \fixp\ both for the case of a
disk with two open string punctures and a sphere with two closed
string punctures.  Since the classical
open string vertices appear at order $\hbar^0$ we  include the
open string BRST kinetic term at order $\hbar^0$ as well. It is then 
natural to have the closed string BRST kinetic term appear at the
same order of $\hbar$ since it is only the total BRST operator in the conformal
theory that defines the boundary operator $\partial$ at the geometrical level.
\foot{It is of course possible to alter the powers of $\hbar$ by 
$\hbar$ dependent scalings of the string fields.}

\noindent
We now list the moduli spaces that appear for the first few orders
of $\hbar$.

\noindent
(i) $\hbar^0$:~ the disk ($b=1$)
with $m\geq 3$ open string punctures.  

\noindent
(ii) $\hbar^{1/2}$:~ the sphere with three closed
string punctures, and the disk with one closed string puncture 
and $m\geq 0$ open string punctures.

\noindent
(iii) $\hbar^1$:~ the sphere with four closed string
punctures, the torus without punctures,
the disk with two closed string punctures and $m\geq 0$
open string punctures, and the annulus 
with $m\geq 0$ open string punctures split in all possible ways on
the two boundary components.  

\noindent
(iv) $\hbar^{3/2}$:~ the sphere with five closed string
punctures, the torus with one closed string puncture,
the disk with three closed string punctures and $m\geq 0$
open string punctures, and the annulus 
with one closed string puncture and
$m\geq 0$ open string punctures split in all possible ways on
the two boundary components.

Note that one can redefine the string
fields and eliminate a separate dependence on $\hbar$ and $\kappa$. 
Indeed it follows from \sumdv\ that (schematically) 
$$\eqalign{
{1\over \hbar} 
\, \V\, &\sim  \sum\,\,\hbar{}^{-\ov\chi + \ov p -1}  \, 
\kappa{}^{-2\ov\chi}
\, \V^{\, g,n}_{\,b,m}\,\cr
& \sim  \sum\,\,(\hbar \kappa^2){}^{-\ov\chi + \ov p -1}  \, 
\Bigl[\kappa{}^{\, n+m}
\, \V^{\, g,n}_{\, b,m} \Bigr] \,, \cr }  \eqn\sumx$$
where use was made of \newfct. It follows from the above equation that
upon a simple  coupling constant rescaling the terms in brackets can
be thought as the new string vertices, and $S/\hbar$ only depends
on $\hbar\kappa^2$.

\section{Sub-recursion relations}

The recursion relations \recmot\ relate all of the string vertices of
the open-closed string theory. All relevant string vertices appear in $\V$.
It is of interest to isolate sub-families of string vertices related
by simpler recursion relations.

\noindent
\underbar{Closed string vertices}
Closed string vertices, namely, surfaces
with no boundaries define a subfamily because
it is not possible to obtain  surfaces without boundaries
by sewing operations on surfaces with boundaries (sewing of open string
punctures glues only pieces of boundary components). 
If we define
$$\V^c = \sum_{g,n} \hbar^p \kappa^q \V_{g,n}\,,\eqn\csvsf$$
we get the recursion relation 
$$\partial \V^c + \half \{ \V^c \,, \V^c \} + \hbar\Delta \V^c = 0 \,,
\eqn\recfone$$ 
for the closed string sub-family. 
The values of $p$ and $q$ need not be those given earlier, since those
were determined by conditions on the three open string vertex. If one
deals with closed strings only one requires that $p=0$ and $q=1$ for
the three punctured sphere, giving  $p=g$, $q= -\ov\chi = 2g+n-2$. Note
that the classical closed string vertices, defined by only summing over $g=0$
in the above, are also a subfamily of vertices.

\noindent
\underbar{Disks with open string punctures} 
These are the vertices that define classical open string field theory.
We define
$$\V_0 = \sum_{m\geq 3}  \V^{\,0,0}_{1,m}  \,, \eqn\defsec$$
dropping, for convenience all factors of $\hbar$ and $\kappa$.  The zero 
subscript denotes zero number of closed strings. 
We must  explain why this is a subfamily. The idea is simple, surfaces
outside this family cannot produce surfaces in the family by sewing operations.
This can be established at the same time as we 
 find the recursion relation satisfied by $\V_0$ by inspection of  Fig.~4.
Here we argue  that only the first term in the right hand side is relevant.
The second configuration
is not relevant because in order to get a surface with one boundary 
component we would need a surface with $b=0$
and then open string sewing is impossible. The third configuration
is also impossible since we would need a string vertex with $g=-1$.
The fourth configuration, having closed string tree-like sewing would require
that the boundary component be in one of the vertices only, the other
vertex must be a pure closed string vertex. Since closed string vertices
start with $n\geq 3$ we would have two external closed string punctures,
in contradiction with the fact that we must just get a disk open 
string punctures. The fifth configuration is irrelevant because of genus.
Therefore, we have shown that
$$\partial \V_0 + \half \{ \V_0 \,, \V_0 \}  = 0 \,.
\eqn\rectree$$ 
The antibracket here is the complete antibracket, but given the 
absence of closed string punctures it just acts on the open string sector.
Note that the number of open string punctures in $\V_0$ goes, in general,
from three to infinity. Thus, in general we have a fully nonpolynomial
classical open string field theory.  In algebraic terms, the structure
governing this theory is an $A_\infty$ algebra.

\goodbreak
\noindent
\underbar{Disks with open string punctures and one or zero closed strings} 
\nobreak
We now introduce the set of vertices described by disks with one closed
string puncture, and $m\geq 0$ open string punctures. 
We define
$$\V_1 = \sum_{m\geq 0}  \,\V^{\,0,1}_{1,m}  \,, \eqn\defsec$$
We now claim that there are recursion relations involving this family $\V_1$ and
 the family $\V_0$.  Again, inspecting  Fig.~4 and
by essentially identical arguments, we argue 
 that only the first term in the right hand side is relevant.
It then follows that 
$$\partial \V_1 +\{ \V_0 \,, \V_1 \}  = 0 \,.
\eqn\recclo$$ 

\goodbreak
\noindent
\underbar{Disks with open string punctures and two or less closed strings}
\nobreak
\noindent
We now introduce the set of vertices described by disks with two closed
string punctures, and $m\geq 0$ open string punctures 
We define
$$\V_2 = \sum_{m\geq 0}  \,\V^{\,0,2}_{1,m}  \,, \eqn\defsecc$$
This time, inspection of   Fig.~4, reveals that, in addition to the 
first configuration, the vertices $\V_1$ can now be sewn to the three
string vertex $V_3^c$ to give a surface with one boundary and 
two closed string punctures. We therefore have t then follows that 
$$\partial \V_2 + \{\V_0, \V_2\} + \half \{\V_1 , \V_1 \}_o + 
\{\V_1, \V_3^c\}= 0 \,. \eqn\rectwo$$
It is fairly clear that we could continue in this fashion adding closed 
string punctures one at a time. If we define $\V_N$ as disks with 
$N$ closed string punctures, their recursion relations will involve
the other $\V_n$'s  with $n<N$, and all classical closed string vertices
up to  $\V^c_{N+1}$.  Since we will not have any explicit use for the particular 
families  $\V_{N>2}$, we now consider all of the families $\V_N$ put together.

\goodbreak
\noindent
\underbar{Disks with all numbers of open and closed string punctures}
We now introduce the set of vertices $\ov\V$ that includes the families
$\V_0, \V_1, \V_2$ introduced earlier along with all their higher counterparts.
$$\ov\V = \sum_{n\geq 0}  \,\V_n \,, \eqn\defsecc$$
This time, inspection of   Fig.~4, reveals that,
$$\partial\ov \V + \half \{\ov \V, \ov\V\}_o + \{\ov\V, \V^c \} = 0 \,. \eqn\reopct$$
Note that the second term is the antibracket on the open string sector only.
Had we used the closed sting antibracket we would get surfaces with two
boundary components.

\bigskip
There are other subfamilies, but we have found no explicit use for them. 
A curious one is the family $\widetilde \V$ that includes the set of all genus zero vertices having
all numbers of boundary components and all numbers of open and
closed string punctures. The vertex $\widetilde\V$ satisfies an equation 
of the type 
$$\partial \widetilde\V + \half \{ \widetilde\V , \widetilde\V \} + \Delta'_o \widetilde \V = 0\,, 
\eqn\nouse$$
where $\Delta_o'$ denotes the open string delta restricted to the case when it acts
on two punctures that are on the same boundary component. This must be the case since
the operation of sewing two open string punctures on different boundary components
of a surface increases its genus.

\chapter{Minimal area string diagrams and low dimension vertices}

In order to write a open-closed string field theory we
have to determine the open-closed string vertices. In other
words we have to find $\V$. This involves specifying, for each moduli
space included in \sumv, the region of the moduli space that
must be thought as a string vertex. Throughout this region we must know 
how to specify local coordinates on every puncture.

The purpose of the present section is to show explicitly how to do this for
the relevant low dimension moduli spaces, and to explain how the 
procedure can be carried out in general. The strategy was summarized in
Ref.[\zwiebachlong]; one defines the string
diagram associated to every surface, and uses this string diagram to
decide whether or not the surface in question belongs to the string vertex. 
A string diagram corresponding to a punctured
Riemann surface $R$, is the surface $R$ equipped with local 
complex coordinates at the punctures.  Since suitable (Weyl)
metrics on Riemann surfaces can be used to define local coordinates
around the punctures [\sonodazw,\zwiebachqcs],
the string diagram for the surface $R$ can be thought  as $R$
equipped with a suitable metric $\rho$.
 
For open-closed string theory we propose again to use minimal
area metrics, this time requiring that open curves be
greater than $\pi$ (to get factorization in open string
channels) and that closed curves be longer than $2\pi$
(in order to get factorization in closed string channels). 
\medskip
{\noindent\narrower\sl{\bf 
Minimal Area Problem for Open-Closed String Theory}:\quad 
Given a genus $g$ Riemann surface $R$ with $b$ boundaries, $m$ punctures
on the boundaries and $n$ punctures in the interior ($g,b,n,m \geq 0$)
the string diagram is defined by 
the metric $\rho$ of minimal (reduced) area under the condition that the 
length of any nontrivial open curve in $R$, with endpoints at 
the boundaries be greater or equal to $\pi$ and that the length of any 
nontrivial closed curve be greater or equal to $2\pi$.\smallskip}
\medskip
As in our earlier work, since the area is infinite when there are
punctures, one must use the reduced area, which is a regularized
area obtained by subtracting the leading logarithmic divergence
[\zwiebach].  All relevant properties of area hold for the reduced area.
Open curves are curves whose endpoints lie on boundary components.

 This minimal area problem produces suitable
metrics. By ``suitable" we have two properties in mind. First, it allows
us to define local coordinates at the punctures. This is because
the metric in the neighborhood of a closed string puncture is that
of a flat cylinder of circumference $2\pi$, and 
the metric in the neighborhood of an open string puncture is that
of a flat strip of width $2\pi$. Both the cylinder and the strips have
well defined ends (where they meet the rest of the surface) and thus define
maximal cylinders (disks) and maximal strips (half-disks). 
These are helpful to define the
local coordinates around the punctures, as will be explained shortly. 

The second property is that
gluing of surfaces equipped with minimal area metrics must give
surfaces equipped with minimal area metrics. This
requires that the definition of local coordinates be done with
some care in order that the sewing procedure will not introduce
short nontrival curves. The maximal domains around the punctures 
cannot be used. Somewhat smaller canonical domains will be used.
The canonical closed string cylinder (or disk) is defined to be bounded
by the closed
geodesic located a distance $\pi$ from the end of the maximal cylinder. If 
we remove the canonical cylinder we are leaving
a ``stub" of length $\pi$ attached to the surface. 
The canonical strip (or half-disk)
is defined  to be bounded by the open
geodesic located a distance $\pi$ from the end of the maximal strip.
If we remove the canonical strip we are thus leaving
a ``strip" of length $\pi$ attached to the surface (see Fig.~5). 
Upon sewing, the left-over stubs
and short strips ensure that 
we cannot generate nontrivial closed curves shorter than $2\pi$ nor
nontrivial open curves shorter than $\pi$.

\Figure{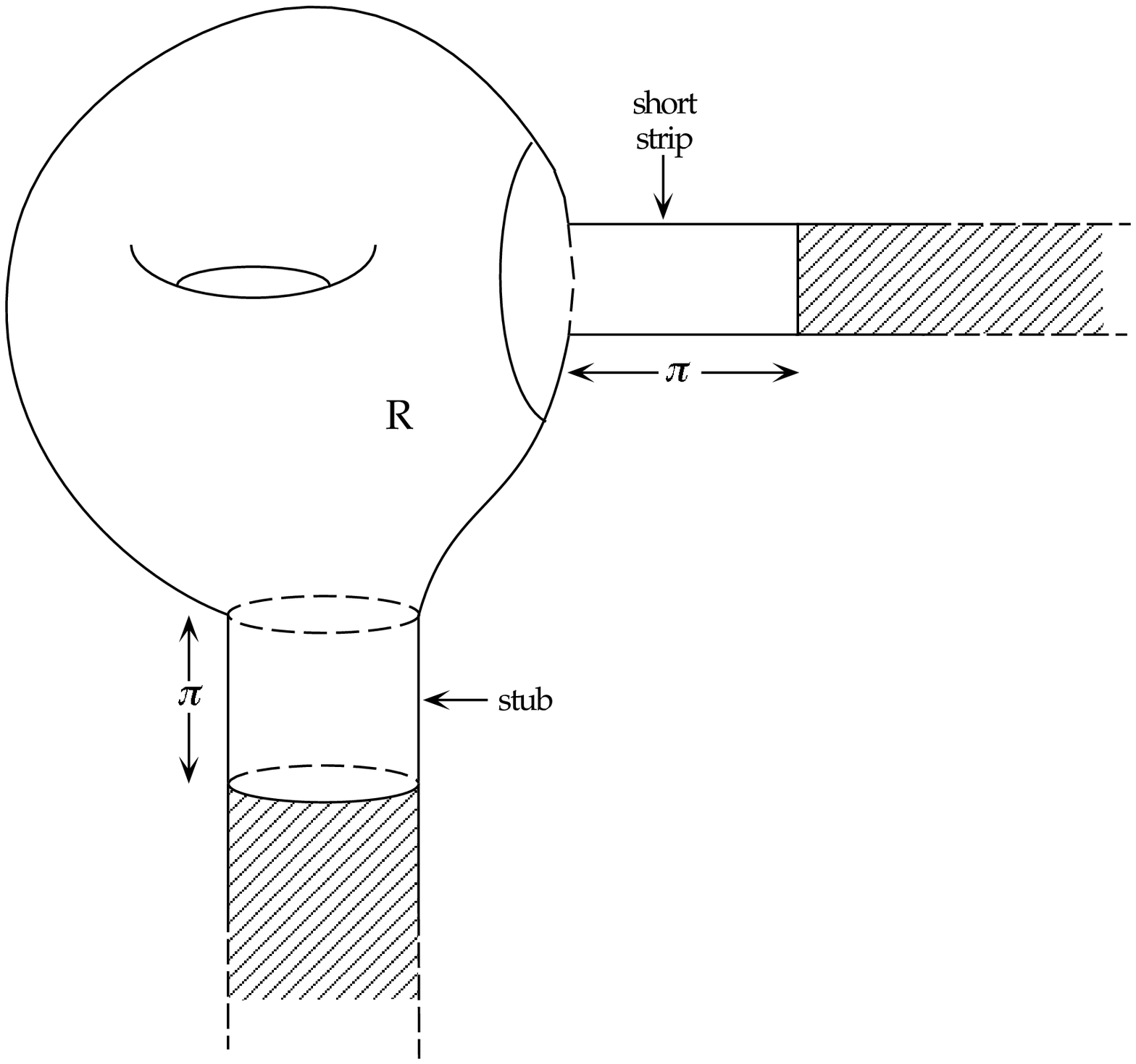 scaled 500}{Figure 5.~ A surface $R$  equipped with a minimal
area metric. The boundaries of the maximal cylinder 
and of  the maximal strip   
are indicated as dotted lines. The canonical cylinder  and the canonical strip
are shown dashed.  They do not coincide with the maximal cylinder
and the maximal disk, but rather differ by  a length $\pi$ cylindrical 
 ``stub" attached to
the surface, and a length $\pi$ short ``strip" attached to the surface. }

The simplest string diagram is the open-closed string 
transition corresponding to a disk with one closed string puncture
and one open string puncture. This string diagram is 
illustrated in Fig.~6(a).
It is very different from the one used
in the light-cone, where an open string just closes up when the
two endpoints get close to each other. Here an open string of
length $\pi$ travels until all of the sudden an extra segment
of length $\pi$ appears and makes up a closed string of length
$2\pi$.  It is simple to prove that this is a minimal area metric.
Imagine cutting the surface at the open string geodesic where
the open string turns into a closed string. This gives us a a semiinfinite
strip and a semiinfinite cylinder.  If there is a metric with less area
on this surface, it must have less area in at least one of these two
pieces. But this is impossible since flat cylinders and flat strips already
minimize area under the length conditions.

\Figure{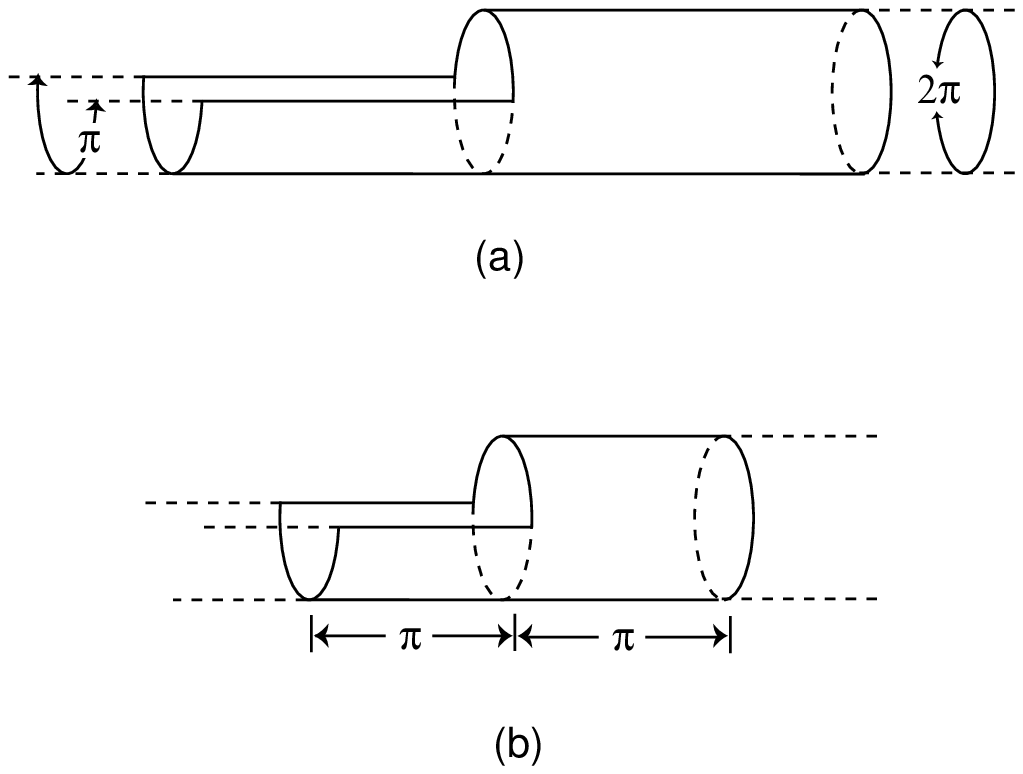 scaled 900}{Figure 6.~ (a)The open-closed minimal area string diagram. 
An open string of length $\pi$
suddenly becomes a closed string of length $2\pi$.
(b) The corresponding string vertex with a stub and a short strip.}

 The open string theory of Witten [\wittenosft] 
is  related to a different minimal area problem.
Its diagrams use the metric of minimal area under the condition that
nontrivial open curves with boundary endpoints be at least
of length $\pi$ [\zwiebachopen,\zwiebachopentwo]. 
This minimal area problem allows for short
nontrivial closed curves, in contrast with the open-closed
problem discussed above. In order to incorporate closed string punctures
into Witten's formulation one  simply declares that the open curves
in the minimal area problem cannot be moved across closed string punctures
[\zwiebachos]. The resulting string diagrams are not manifestly factorizable
in the closed string channels.

\goodbreak
\noindent
\underbar{The specification of the string vertex}
\nobreak
In general, the minimal area problem determines the string 
vertex $\V$ as follows. Given a surface $R$, we find its minimal
area metric. Then $R\in \V$ ($R$ belongs to the string vertex)
if there is no internal propagator in the string
diagram. Namely, if we cannot find an internal flat cylinder or
an internal flat strip of length greater than or equal to $2\pi$. 
The logic is simple [\zwiebachlong]. Since string vertices have
stubs and short strips, whenever we form a Feynman graph by sewing
we must generate either cylinders or strips longer than $2\pi$. If such
cylinders or strips cannot be found in a string diagram, 
the string diagram in question
cannot have arisen from a Feynman graph, and must therefore be included
in the string vertex.

\section{Classical open string theories}
For surfaces corresponding to open string tree 
amplitudes (disks with punctures on the boundary), both 
the open-closed minimal area problem of this paper,
and the pure open minimal
area problem alluded above yield the same minimal area metrics. This
is because such surfaces have no nontrivial closed curves. For the case
of loop amplitudes the two minimal area problems give
different metrics. 

Even though the metrics are the same for
the case of disks with open string punctures, the 
string  vertices {\it are not the same}. In [\wittenosft] there is
only a three open string vertex corresponding to the disk with
three open string punctures. In classical open-closed
string field theory there will be open string vertices for
corresponding to disks with $m$ open string punctures, for all
values of $m\geq 3$. This is due to the inclusion of length $\pi$ open
strips on the legs of the string vertices. 
The higher classical string vertices can be obtained as the 
string diagrams built using  Witten type  three-string vertices  and
open string propagators shorter than $2\pi$. 

If we are only concerned about the classical theory of open strings
the open strips attached to the vertices can have any arbitrary length
$l$. The family of vertices $\V_0 (l)$ introduced earlier would depend on
the parameter $l$. For $l=0$ we find  the Witten theory, including
only a three punctured disk, while for any $l\not= 0$ we get  moduli spaces
of disks with all  numbers of open string punctures.  in particular
for $l=\pi$ we get the classical open-closed string theory.  For all values
of $l$ the recursion relations $\partial \V_0 (l) + \half \{ \V_0 (l) , \V_0 (l) \} = 0$ hold.
It is clear that the parameter $l$ defines a deformation of open string vertices interpolating
from the Witten theory to the open-closed string theory of this paper. As shown by
Hata and the author in the context of closed strings [\hatazwiebach], deformations of string vertices preserving the recursion relations give rise to string field theories related by
canonical transformations of the antibracket.  As elaborated by Gaberdiel and
the author [\gaberdielzwiebach], this means that the family of  $A_\infty$ homotopy algebras ${\bf a} (l)$ underlying these $l$-dependent string field theories 
are all homotopy equivalent in the same vector space.

\section{Extracting the Dimension Zero Vertices}

Out of the dimension zero vertices (listed in \S2.1) there are six that do not
appear in the string vertex $\V$.  These are the zero, one and twice punctured sphere,
and the disk with zero, one or two open string punctures.
The above two-punctured surfaces are used as building blocks for
the symplectic forms. The above once-punctured surfaces would be
relevant in formulations around non-conformal backgrounds. The  surfaces
with no punctures would only introduce constants (see, section 9, however).

Next are the three
punctured sphere and the disk with three open punctures.
Those are relevant vertices and are taken to correspond to
be the symmetric closed string vertex and the symmetric
open string vertex [\wittenosft] since they solve the minimal area problem.
For the closed string vertex we must include  a stub of length
$\pi$ in each of the three cylinders,
and for the open string vertex we must include  a short strip
of length $\pi$ on each of the three legs (Fig.~7) 

\Figure{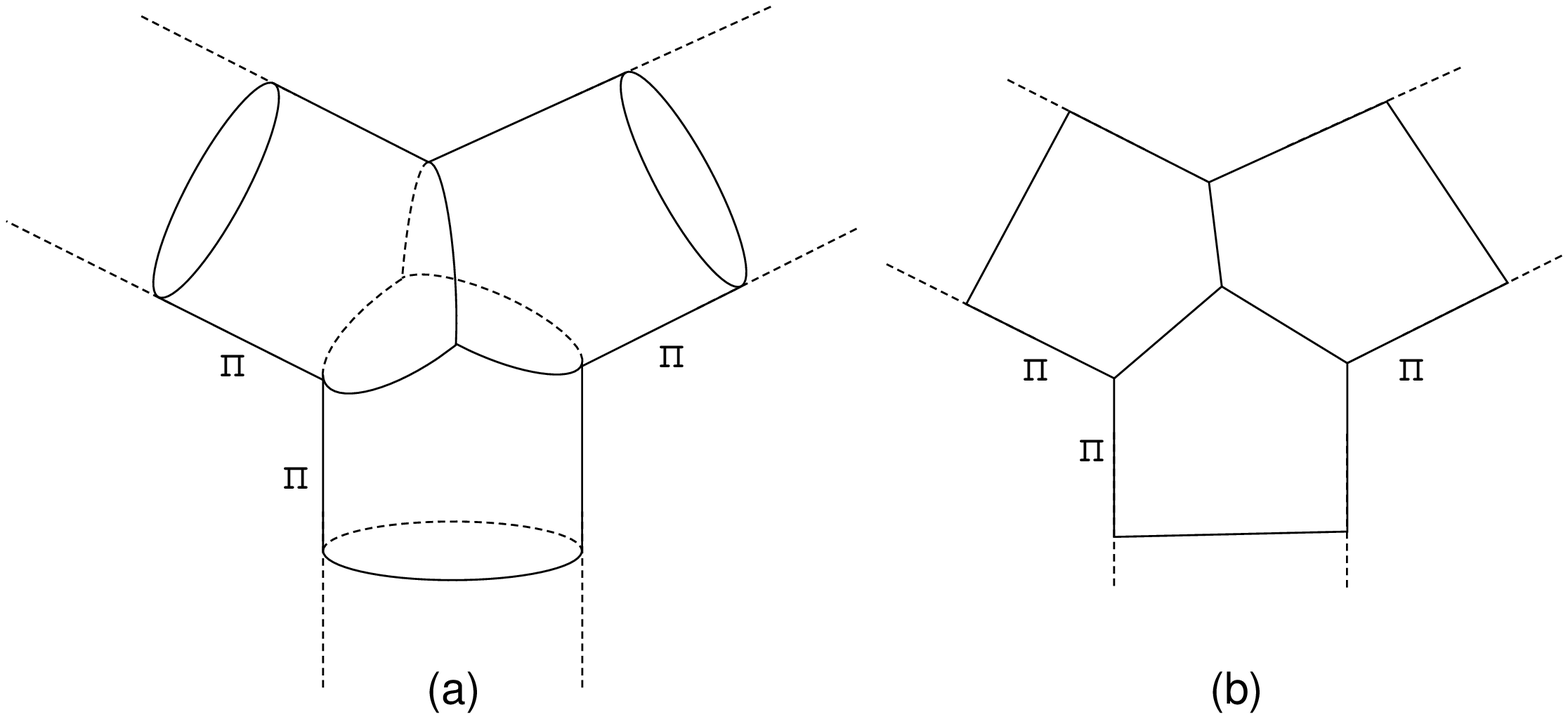 scaled 500}{Figure 7.~ (a) The three closed string vertex with its stubs
of length $\pi$ included. (b) The open string vertex with its short
strips of length $\pi$ included.}

Next at dimension zero is the disk with one 
open and one closed puncture, corresponding to the open-closed 
string diagram considered before in Figure 6(a). The vertex, shown in
Fig.~6(b) includes a stub and a  short strip. Finally, we have a disk with 
one closed string puncture only. This surface represents a closed
string that just stops. In our minimal area framework it
is a semiinfinite cylinder of circumference $2\pi$
(Fig.~8(a)). As a vertex,  it looks like a short cylinder
with two boundaries, one to be connected to a propagator and
the other left open (Fig.~8(b)). This surface has a conformal 
Killing vector.

\Figure{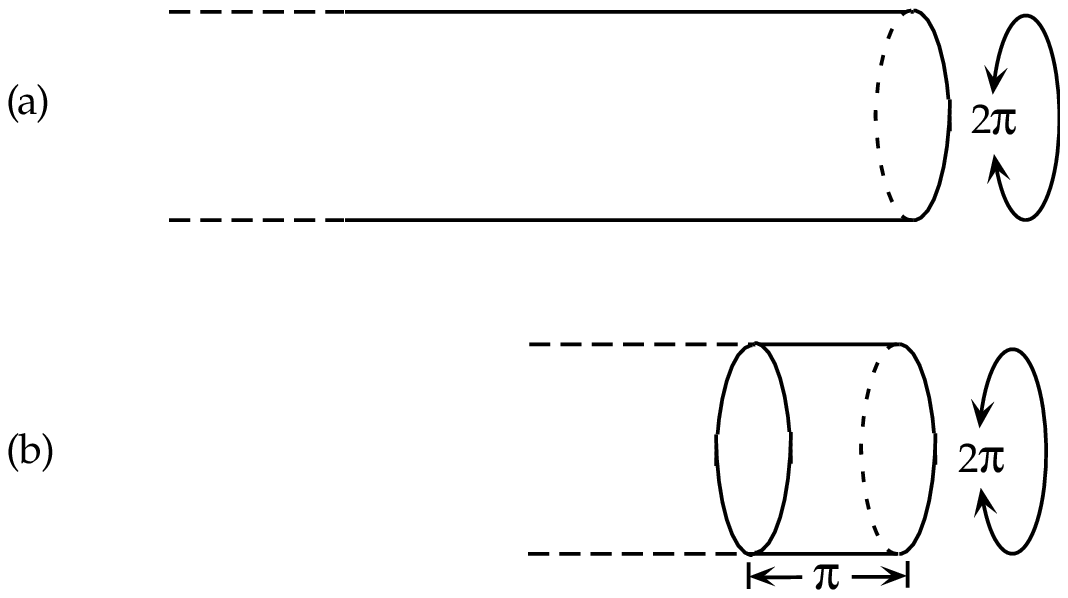 scaled 900}{Figure 8.~ (a) This is the string diagram for a disk with
a closed string puncture. It represents an incoming closed string that simply 
stops. (b) As usual, we must leave a stub of length $2\pi$
in the corresponding string vertex.}

\section{Extracting the Dimension One Vertices}

As listed in \S2.1 there are four moduli spaces that must 
be considered. We examine each of them  to determine both the minimal
area string diagrams 
and  to extract the string vertices.

\noindent
\underbar{The  disk with four open string punctures}~  The 
boundary of a  moduli space  of  four punctured disks 
arises from  two three-string 
open vertices sewn together (see Fig.~4).  For a given cyclic labelling of the punctures
there are four superficially different ways to assign cyclically the labels in the sewn
diagram.  The four configurations break up into two groups of  identical
configurations, the factor of one-half in the geometrical equation showing  
that  the boundary of the moduli space in question is given by two
sewing configurations, one with a plus sign and one with minus sign (recall the
cyclic factor is minus one to the number of punctures minus one).  This is the
familiar fact that an open string four vertex must interpolate from an $s$-channel
to a $t$-channel configuration.  
Since the three string vertex has a short
strip of length $\pi$ attached, the boundaries of the moduli space of
four punctured disks are minimal area metrics with 
internal strips of length $2\pi$.  The four string
vertex that we need  includes all  four punctured disks  whose
minimal area metrics have an internal strip of length  less than or equal 
to $2\pi$.  The external legs are amputated leaving strips of length $\pi$. 

\noindent
\underbar{The disk with one closed string puncture and 
two open string punctures} 

\noindent
This moduli space corresponds to an open-open-closed
interaction, and will be discussed in detail in section 6.3. 
Since the discussion in that section will be based on the 
associative open string vertex, the open-open-closed vertex of the
open-closed theory also  includes the region of 
moduli space that arises when the closed-open vertex is joined
to the three-open-string vertex, and we let the  intermediate strip shrink
from the initial length of $2\pi$ down to zero.

\Figure{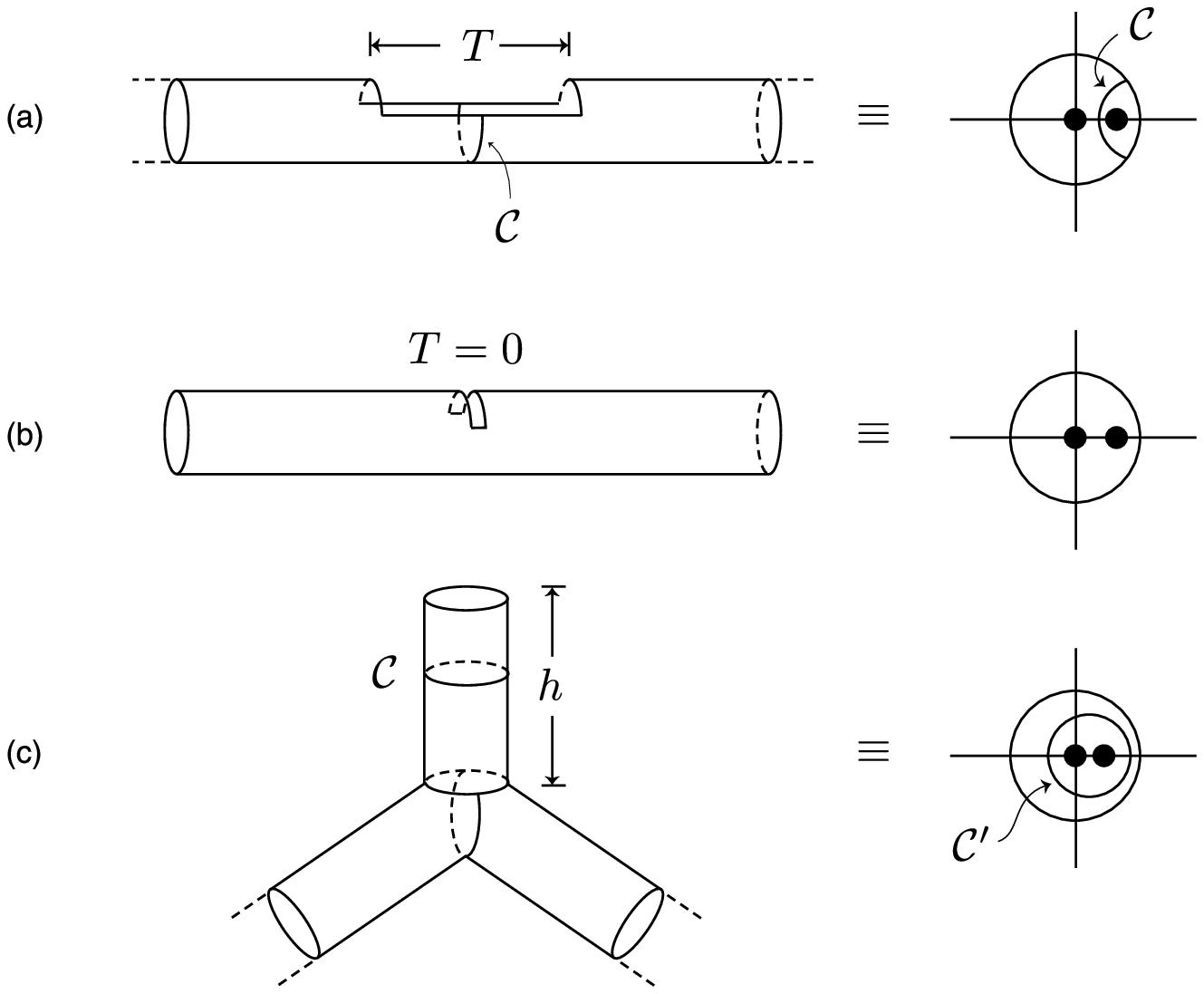}{Figure 9. Minimal area string diagrams for a disk with two closed
string punctures. (a) A closed string turns into an open string that
propagates  for time $T$ and then turns back into a closed string.
When $T\to \infty$ we get open string poles. (b) The above configuration
when $T=0$. We have a slit on the infinite cylinder. (c) A stub of
length $h$ emerges from the slit. As $h\to \infty$ we get closed
strings going into the vacuum.}

\noindent
\underbar{Disk with two closed string punctures}
The moduli space
can be represented by taking a unit disk with one of the
closed strings in the center and the other in the positive 
horizontal axis. The position of the second puncture is the
modular parameter. Consider a closed string turning into an 
open string and turning back into a closed string, as shown 
in Fig.~9(a). The intermediate time is denoted by $T$. It is clear
that for any value of $T$ this is a minimal area surface, since 
its double is a closed string diagram, a four closed string
scattering amplitude with an intermediate propagator. 
When $T\to \infty$ we find open string poles. This 
is due to the presence of short noncontractible open curves in the disk representation.
In this representation one of the punctures is approaching the
boundary and the short curve is ${\cal C}$ (Fig.~9(a)).
When $T$ becomes zero one is
left with a cylinder with a slit of perimeter $2\pi$ (Fig.~9(b)). But
this cannot be the end of the story. There is nothing singular
about this surface, and therefore we are missing the region 
where  the two closed string punctures are very close to
each other in the disk representation. When this happens 
a new closed string appears, the noncontractible 
curve surrounding the two punctures, this closed string is
homotopic to the boundary of the disk. Indeed, in the minimal area string 
diagrams when $T$ becomes zero a cylindrical stub begins to grow
out of the slit (Fig.~9(c)). The height $h$ of the stub goes from zero
to infinity supplying the remaining region of moduli space.
The curve ${\cal C}$ represents the new closed string that
is going into the vacuum. It is clear that if the 
boundary is mapped to the unit disk
the punctures are becoming close to each other as $h\to \infty$.
This region of moduli space can also be pictured as a two
punctured sphere with a boundary that is becoming smaller 
and smaller. This viewpoint arises if we demand, in the
disk representation, that the two external closed string 
punctures be fixed. Then
the boundary must grow until it becomes a tiny circle around
infinity. As the boundary shrinks we are obtaining a three
punctured sphere. Indeed, as $h\to \infty$,  the minimal
area string diagram becomes  the
standard minimal area diagram representing a three punctured sphere.

Clearly, the region $T \in [2\pi ,\infty \}$ is
produced by the Feynman graph using two open-closed 
string vertices and an open string propagator. Moreover, 
the region $h\geq 2\pi$ is produced by a graph containing
the three-closed-string vertex, a propagator and the
closed-boundary vertex. The missing region $(T\leq 2\pi )
\cup (h\leq 2\pi )$ must be supplied by the closed-closed-boundary
vertex. As usual, the vertex is defined by the set of surfaces,
each having the external strips and cylinders amputated down to
length $\pi$.

\Figure{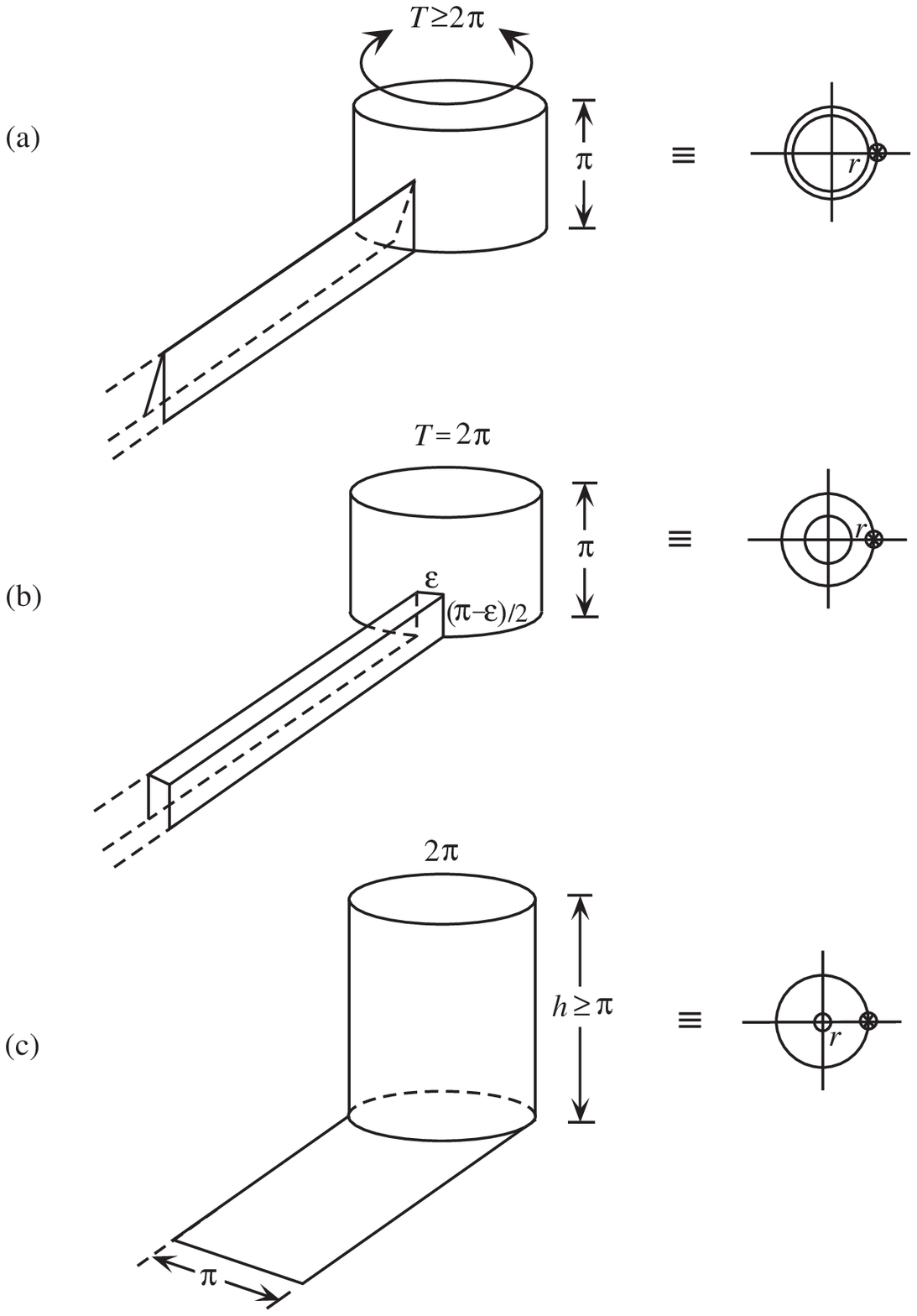 scaled 900}{Figure10. Minimal area string diagrams for an annulus
with an open string puncture. (a) When $T\to \infty$ we get open string
poles. The length $T$ cannot be shorter than $2\pi$, since this
would violate the length condition on nontrivial closed curves.
(b) When $T=2\pi$ the slit gets deformed into a rectangular hole
of width $\epsilon$ and height $(\pi - \epsilon )/2$. When
$\epsilon = \pi$ the configuration turns into that of (c), with
$h=\pi$. When $h\to \infty$, we get closed strings going into the vacuum.}

\goodbreak
\noindent
\underbar{Annulus with one open string puncture}

The outer boundary of the
annulus is set to be the  $|z|=1$ boundary of the disk, the open string puncture
is at $z=1$, and the modular parameter is chosen
to be the radius $r$ of the inner boundary of the
annulus (see Fig.10) . When $r\to 1$ one has very short open curves
going from one boundary to the other, showing the 
presence of open string poles. The relevant string
diagram in this region is given by the ``tadpole"
 graph shown in Fig.~10(a), where the length
$T$ of the intermediate strip is very big, in particular
much  longer than
$2\pi$. This is a solution of the 
minimal area problem because its double is a 
closed string diagram, a one loop tadpole with
a propagator of length $T$.

In the Witten formulation of open string theory one
lets $T$ go all the way to zero where  a
singular configuration is encountered (Fig.11).  The
Fock space representation of this configuration
involves divergent quantities, and gives a 
potential violation of the BV master equation [\thornpr].
\footnote{*}{ An early discussion arguing
that the term can be set to zero can be found
in [\bogojevic]. Reference [\maeno] has shown that the BV master
equation is satisfied in the midpoint formalism
of open string theory. The midpoint
formulation of open string theory [18] also
requires careful regularization.} 

\Figure{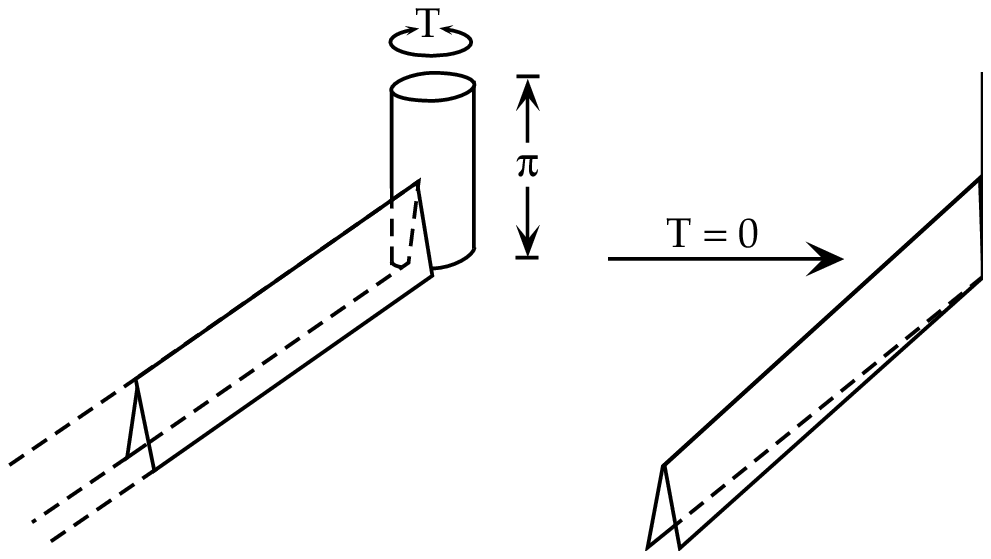 scaled 900}{Figure 11. An open string one loop tadpole. 
The parameter $T$ is the length of the internal
propagator. As $T\to 0$ we obtain a singular limit. }

In open-closed theory, however, we
do not  run into problems since $T$ cannot
shrink below $2\pi$. In fact,
due to the strips of length $\pi$ in the open 
string vertex, this endpoint is a collapsed tadpole. 
While we do not know the explicit
minimal area metrics,  what happens after that must be fairly 
close to the following: the slit must be deformed into a 
rectangular hole of width $\epsilon$ and height 
$(\pi - \epsilon)/2$, as shown in Fig.~10(b), 
with $\epsilon \in [0,\pi ]$. When $\epsilon =\pi$
we get  a minimal area metric matching smoothly with the $h=\pi$ 
configuration
of Fig.~10(c). Then $h$ begins to grow.  When
$h= 2\pi$, the string diagram corresponds to a Feynman
graph built with an open-closed vertex and a closed-boundary vertex in
the limit when the propagator collapses. This
graph provides the remaining part of the moduli
space. The region  $h\to \infty$ corresponds
to closed strings going into the vacuum, or $r\to 0$.
The elementary vertex must therefore comprise
the region 
$\{ 0 \leq \epsilon \leq \pi \} \cup  
\{ \pi \leq h \leq 2\pi \}$.

\chapter{The open-closed master action}

The aim of the present section is to describe the construction of
the open-closed string field action, and to explain why our earlier
work on the string vertices implies that the master equation is satisfied.

The string field action is a function on the full vector space  $\H$ of the
open-closed theory.  This vector space is the direct sum $\H = \H_o \oplus \H_c$
where $\H_o$ denotes the open string state space, and $\H_c$ denotes the
closed string state space. As is well known (see, for example {\zwiebachlong])
$\H_c$ is defined as the set of states of the matter-ghost CFT that
annihilated both by $b_0^-$ and $L_0^-$.
There is no such constraint for $\H_o$, which comprises all states of the boundary
ghost-matter CFT.  

The complete space $\H$ must be equipped with a invertible odd bilinear
form, the symplectic structure of BV quantization.  The symplectic structure we introduce
is  diagonal;  it couples vectors in $\H_c$ to vectors in $\H_c$, and vectors
in $\H_o$ to vectors in $\H_o$.  The bilinear form must also be odd, and therefore
must couple odd vectors to even vectors. 
Recall that in the closed string theory [\zwiebachlong] 
the symplectic  form was described as a bra  $\bra{\omega_{12}^{~c}} \in \H^*_c
\otimes \H^*_c$.
This bra is used to define 
$\langle A, B\rangle_c \equiv \bra{\omega_{12}^{~c}} A\rangle_1 \ket{B}_2$. 
Given two closed string states, the bilinear form $\langle \cdot , \cdot \rangle_c$
gives  the correlator of the two states and a  $c_0^-$ ghost insertion on a  canonical two-punctured.  Since the sum of ghost numbers in a closed string correlator
is six, this insertion guarantees that the bilinear form  couples vectors
of odd ghost number to vectors of even ghost number, and as a consequence
odd vectors to even vectors.  Moreover, the ghost insertion resulted in the
fact that $\bra{\omega_{12}^{~c}} = - \bra{\omega_{21}^{~c}} $.  This property implies
that the string field should be Grassmann even in order for the obvious 
kinetic term $\langle \Psi , Q_c\Psi \rangle$ not to vanish.

 For open strings the correlator of two states on a disk requires that the 
sum of ghost numbers be equal to three. We can therefore naturally
obtain an odd  symplectic form  $\langle \cdot , \cdot \rangle_o$  by defining it to 
be give correlator of two states, without extra insertions, on the canonical
two-punctured disk.  As a Riemann surface, the 
canonical two-punctured disk  is  symmetric under the 
exchange of the punctures.  We may then be tempted to conclude that
for open strings $\bra{\omega_{12}^{~o}} = +\bra{\omega_{21}^{~o}} $.
If that is the case, the naive open string kinetic term $\langle \Phi , Q_o \Phi \rangle$
vanishes unless the string field is odd.  This is in fact the  choice in the formulation of Ref.[\wittenosft].  It seems more convenient, however, 
to have an even open string field, since this allows us to treat the closed 
 and the open string field with the same set of conventions. In the odd string 
field convention, the moduli spaces entering into the string action would 
have to have nontrivial
degrees according to the number of open strings they couple, rather than degree
zero, as we described in an earlier section.  The distinction between having
an odd or an even string field is at any rate a matter of convention, and we
will work with the even string field convention.

In the odd
string field convention, the typical classical bosonic open string field  $c_1 V \ket{0}$ is odd, 
and the in-vacuum $\ket{0}$ is declared an even  vector.\foot{While our remarks
are all made explicitly in the context of bosonic strings, they are also
applicable to the case of NS superstrings.}  
It follows from $\bra{0}c_{-1}c_o c_1 \ket{0} = 1$ that the out-vacuum $\bra{0}$
should be declared odd.  In the conventions to be used here
we  take $\ket{0}$ to be odd and $\bra{0}$ to be even. The string field will be even,
and we should expect the symplectic form to be antisymmetric under
the exchange of its labels $\bra{\omega_{12}^{~o}} 
= - \bra{\omega_{21}^{~o}}$, as this is required for the canonical kinetic term
$\langle \Phi , Q_o \Phi\rangle_o$ not to vanish when the string field is even.
We can examine the exchange property explicitly for the case of open bosonic 
strings. In this case the Fock space representation of the bra is given by [\thornpr]
$$\eqalign
{\bra{\omega^{~o}_{12}} &= \int  {dp \over (2\pi)^d} \,\,  {}_1\bra{p} c_{-1}^{(1)} \,\,\cdot
\,\,  {}_2\bra{-p} c_{-1}^{(2)}  \,\,\cdot \,  (c_0^{(1)}  + c_0^{(2)}  ) \, \cr
& \quad\,\, \cdot \, \exp \Bigl[ - \sum_{n=1}^\infty
(-)^n (\alpha_n^{(1)}\cdot  \alpha_n^{(2)}  + c_n^{(2)} b_n^{(1)} 
+ c_n^{(1)} b_n^{(2)} \, )  \Bigr] \,. \cr} \eqn\osymp$$
Since in our convention $\bra{0}$ is even, the bra $\bra{p}$ is also even, and
$\bra{p} c_{-1}$ is odd.  It follows from examination of the first line of the above
equation that the bra $\bra{\omega^{~o}}$ is indeed antisymmetric under the
exchange of labels.

The open and closed bilinear forms satisfy
$$\eqalign{ \bra{\omega_{12}^{~o} } ( Q_o^{(1)} + Q_o^{(2)} ) &= 0 \, ,\cr
 \bra{\omega_{12}^{~c} } ( Q_c^{(1)} + Q_c^{(2)} ) &= 0\, , \cr}\eqn\brstp$$
and both are invertible, with symmetric inverses $\ket{S^o_{12}} =  \ket{S^o_{21}} $
and  $\ket{S^c_{12}} =  \ket{S^c_{21}} $ satisfying
$$\eqalign{ \bra{\omega_{12}^{~o} }S^o_{23}\rangle = {}_3 {\bf 1}^o_1, \quad
 ( Q_o^{(1)} + Q_o^{(2)} )\ket{S^o_{12}} &= 0 \, ,\cr
\bra{\omega_{12}^{~c} }S^c_{23}\rangle = {}_3 {\bf 1}^c_1, \quad
 ( Q_c^{(1)} + Q_c^{(2)} )\ket{S^c_{12}} &= 0 \, ,\cr}\eqn\brstpi$$
The string fields are defined by the usual expansions
$$\ket{\Phi} = \sum_i \ket{\Phi_i} \phi^i \, , \quad \ket{\Psi} = \sum_i \ket{\Psi_i} \psi^i \, , $$
where $\ket{\Phi_i}$ and  $\ket{\Psi_i} $ are basis vectors for the state spaces
$\H_o$ and $\H_c$ respectively, and  $\phi^i$ and $\psi^i$ denote the target space open
and closed string fields respectively.

For functions on the state space $\H$ we define the BV antibracket and the delta operator
by the expressions 
$$\{ A , B \} = {\partial_r A\over \partial \phi^i} \, \omega_o^{ij}\,
 {\partial_l B\over \partial \phi^j}
\, + \,  {\partial_r A\over \partial \psi^i}\,  
\omega_c^{ij}\,  {\partial_l B\over \partial \psi^j}  \,, \eqn\antibexp$$
$$\Delta A = {1\over 2} (-)^i {\partial_l\over \partial \phi^i} \Bigl( \omega_o^{ij} {\partial_l A
\over \partial \phi^j }  \Bigr) 
+  {1\over 2} (-)^i {\partial_l\over \partial \psi^i} \Bigl( \omega_c^{ij} {\partial_l A
\over \partial \psi^j } \Bigr)\, , \eqn\deltaexp$$
where the matrices $\omega_o^{ij}$ and $\omega_c^{ij}$ are the inverses of the matrices
$\omega^o_{ij}$ and $\omega^c_{ij}$ respectively.  The latter are defined from the
symplectic forms as $\omega^o_{ij} = \langle \Phi_i , \Phi_j \rangle_o$ and 
 $\omega^c_{ij} = \langle \Psi_i , \Psi_j \rangle_c$.

The open-closed string field action $S$ is
a function in $\H$, or in other words it is a function of both the open and closed
string fields $S(\Phi , \Psi)$. Consistent quantization requires that it should satisfy
the BV master equation
$$  \half \{ S , S \}  + \hbar \Delta S= 0 \,. \eqn\bvmast$$
In order to construct an action that manifestly satisfies this equation we must
now consider a map from the complex $\P$ of moduli spaces of bordered surfaces
to the space of functions on $\H$. We define
$$ f (\A^{g,n}_{b,m}  ) =\Bigl[  {1\over n!} {1\over b!} \prod_{k=1}^b {1\over m_k} \Bigr] \, 
\int_{\A}  \bra{\Omega} \underbrace{\ket{\Psi} \cdots  \ket{\Psi} }_{n} 
\prod_{k=1}^b  \underbrace{\ket{\Phi} \cdots  \ket{\Phi} }_{m_k} \eqn\rept$$
where we have included normalization factors $n!$ for the number of closed strings,
$b!$ for the number of boundary components, and $m_k$ for each boundary component.
In the integrand we insert $n$ copies of the closed string field on the $n$ punctures,
and $m_k$ copies of the open string field on the $k$-th boundary component. 
The bra $\bra{\Omega}$ denotes a form on the moduli space of surfaces, its precise
definition for the case of closed strings was given in Refs.[\zwiebachlong, \senzwiebachtwo],
and no doubt a similar explicit discussion could be done for the open string sector.
The main property of the map $f: \P \to C^\infty (\H)$ is that it defines a map of BV
algebras. In analogy to the closed string case discussed explicitly in [\senzwiebachtwo]
we now should have
$$\eqalign{  f (  \{ \A , \B \} ) &= - \{  f(\A) , f (\B) \}  \,, \cr
  \Delta ( f(\A)) & = - f ( \Delta \A) ) \, . \cr} \eqn\hombv$$
With a little abuse of notation we will denote by $Q_o$ the function in $\H$ that
defines the open string kinetic term, namely $\half \langle \Phi , Q_o \Phi\rangle_o$,
and by $Q_c$ the function in $\H$ that
defines the closed string kinetic term, namely $\half \langle \Psi , Q_c \Psi\rangle_c$.
Finally, we let $Q= Q_o + Q_c$. We then have
$$\{ Q , f(\A) \} = - f(\partial \A)  \, , \eqn\qpart$$
as expected from the analogous closed string relation [\senzwiebachtwo].

We can now simply set
$$S = Q + f (\V) \, , \eqn\theaction$$
and it follows immediately from  \hombv\ and \qpart\  that
$$  \half \{ S , S \}  + \hbar \Delta S = - f \Bigl (\partial\V  + \half \{ \V , \V \} +  \hbar  \Delta \V\Bigr) \, \eqn\getnh$$
where use was made of $\Delta Q = 0$ [\zwiebachlong]. It  is now clear that the
master equation \recmot\ satisfied by the moduli space $\V$ guarantees that the
master action satisfies the BV master equation.

\chapter{Vertices coupling open and closed strings via a disk}

In this section we will study in detail the simplest couplings
of open strings to closed strings, namely, couplings via a disk.
The Riemann surfaces are therefore disks with punctures on the
boundary, representing the open strings, and punctures in the
interior, representing the closed strings. We will give the
complete specification of the minimal area metric for these
surfaces. This will give us insight into the way the minimal area
problem works, and in addition it will make our discussion very
concrete. In all our analysis we will have in mind the classical
open string field theory of Witten. 
 
These vertices are of particular interest since they enter in the
construction of  a global closed string symmetry of open
string field theory (sect.7), and in the  construction of  an open
string field theory in a nontrivial closed string background (sect.8).

We will begin by discussing some algebraic aspects of the coupling of a closed
string to a boundary, and then some algebraic properties of the   
open-closed vertex, in particular maps of cohomologies 
(the string diagrams corresponding to these two vertices were described earlier.).
We then turn to the open-open-closed vertex, which is also examined
in detail. It is then possible to consider the case of $M$ open strings
coupling to a single closed string. We then turn to the case of two
closed strings coupling to an open string (this vertex actually
vanishes in the light-cone gauge). After studying very explicitly
this case, we are able to generalize to the case of $M$-open strings
coupling to two closed strings, and then to the case of 
$M$ open strings coupling to $N$ closed strings.

\section{Closed String coupling to a empty boundary} 

The vertex was described before in section 4.1 (Fig.8).
The Fock space state associated to this vertex is called the boundary
state.  If we denote by $\H_c$ the vector space of the closed string field, 
the boundary state $\bra{B}$ should be thought as an element of $\H^*_c$
the dual vector space to $\H_c$.  Here we will only make some remarks
on the alternative ways of describing the boundary state.

We may recall that if $H_{cft}$ denotes the state space of the CFT describing
the closed string, then the subspace $\H_c$
$$\H_c = \{ \ket{\Psi} \in H_{cft} \, \bigl|   b_0^- \ket{\Psi} = 0, \, L_0^- \ket{\Psi} = 0  \} $$
is the vector space associated to the closed string field.  Its dual space
is therefore described as the space of equivalence relations
$$\H^*_c  = \{ \bra{\Psi} \in H^* ; \,\,   \bra{\Psi} \sim \bra{\Psi} + \bra{\Lambda} b_0^- , \, \, 
 \bra{\Psi} \sim \bra{\Psi} + \bra{\Lambda'} L_0^-\, .  \} \eqn\sfed$$
Note that for any state in $\H^*_c$ one can always find a representative that is 
annihilated by $c_0^-$ and $L_0^-$.  Therefore one can choose $\bra{B}$ to
satisfy $\bra{B} c_0^- =0$ and  $\bra{B} L_0^- =0$. On the other hand, strictly
speaking, the fact that the boundary state is annihilated by $Q_c$ just means
$$ \bra{B} Q_c = 0 , \,\, \hbox{on} \, \H^*_c   \Leftrightarrow   \bra{B} \, Q_c = \bra{*} b_0^- \, ,
\eqn\meani$$
for some state $\bra{*}$.   
In general the boundary state $\bra{B}$ can couple to every physical state
in $\H_c$. This can be seen as follows. We can rewrite 
$$\bra{B}\Psi\rangle 
= {}_3\bra{B} \bra{\omega_{12}^c} S_{23}\rangle \ket{\Psi}_1 = \langle B , \Psi \rangle_c\, 
\quad\hbox{with} \quad \ket{B}_1= {}_2\bra{B}S_{12}\rangle \, , \eqn\repdiff$$
 with  $\ket{S_{12}} = b_0^- \ket{R_{12}'}$ the inverse
of the closed string symplectic form.  Note that 
\meani\ implies that $Q_c \ket{B} = 0$.
Since the closed string inner product $\langle \cdot
, \cdot \rangle_c$ is nondegenerate [\zwiebachlong], for every 
physical $\ket{\Psi}$ one can find
some state $\ket{B}$ that gives a nonzero inner product. Typically, the restriction
of the bilinear form to physical states is nondegenerate, and in this case 
the state $\ket{B}$ can be chosen to be physical (see, for example, appendix C of
[\gaberdielzwiebach]).  Since the ghost number of $\ket{\Psi}$ is two, and the 
bilinear form couples states of ghost numbers adding to five,  $\ket{B}$ must be
of ghost number three.  Therefore, {\it the physical content of the boundary
closed string vertex is defined by a state in the cohomology of $Q_c$ at ghost number 
three}.  

In bosonic string theory the physical cohomology sits at ghost number two, but
a complete copy of this cohomology sits at ghost number three.
At ghost number two, all physical states with the exception of the ghost dilaton,
are of the form $c_1 \bar c_1 \ket{m}$, where $\ket{m}$ denotes state built of matter
operators. For these states, their ghost number three physical counterparts are of the
form $(c_0+ \bar c_0)c_1 \bar c_1 \ket{m}$.  Even for the ghost dilaton $\ket{D_gh} 
= (c_1 c_{-1} - \bar c_1 \bar c_{-1}) \ket{0}$ its ghost number three counterpart
is obtained by multiplication with $c_0 + \bar c_0$. It  is known, however, that
if one extends suitably the complex $\H_c$ to include the zero modes $x$ of the
bosonic coordinates, all cohomology at ghost number three vanishes identically 
[\astashkevich, \belopolskyzw]. 
This reflects the fact that one can remove the one point functions of the 
closed string physical fields by giving space-time dependent expectation values to
background fields.

\Figure{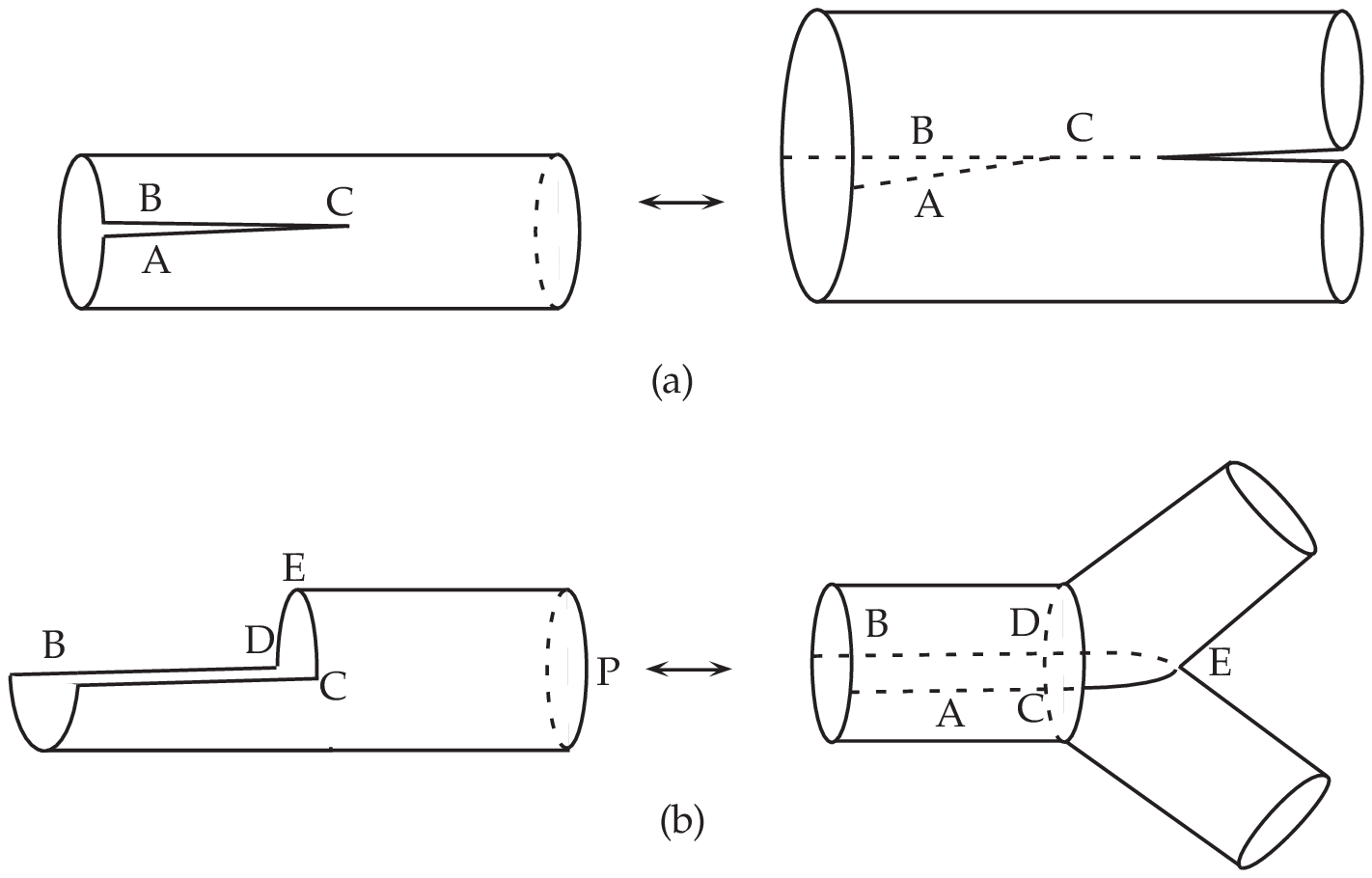 scaled 700}{Figure 12.  (a) An open-closed light-cone vertex, the open
string endpoints lie on $A$ and $B$. At $C$ the endpoints join to make
a closed string. The open-closed light-cone vertex when doubled
gives us the three closed string light cone vertex. (b) The double
of the covariant open-closed vertex gives the standard symmetric
three closed string vertex. }

\section{The Open-Closed String Vertex}

Light-cone field
theory has an open-closed string vertex.
It corresponds to an open string of some fixed length closing
up to make a closed string of the same length [4]. Its double
corresponds to a three-closed string light-cone vertex (Fig.~12). 
 The covariant open closed vertex was examined earlier and is shown 
again in Fig.~12(b). Its double, shown to the right, is the three
closed string diagram. 

 This vertex is denoted in the
operator formalism as the bra $\bra{\V_{oc}} \in \H_o^*\otimes \H^*_c$, 
and the general arguments of [\alvarez] imply that it is annihilated by
the sum of BRST operators:
$$\bra{\V_{oc}} \, (\, Q^o + Q^c\,) = 0.\eqn\xone$$
This vertex couples  open string states $\ket{\Phi}$, to closed
string states $\ket{\Psi}$. Since the surface has no conformal 
Killing vectors the sum of the respective ghost numbers must be three.
\foot{We are assigning zero ghost number to both the 
 SL(2,R) and SL(2,C) vacua.}
The term in the string action is given by $\L_{oc} \sim \vev{\V_{oc} | \Phi }\ket{\Psi}$,
and it should be noted that ghost number allows the coupling
of classical open strings (ghost number one) to classical closed
strings (ghost number two).

The open-closed string vertex naturally defines two maps; one  from $\H_o$ to $\H^*_c$
and the other  from $\H_c$  to $\H_o^*$.  By composing  these maps 
with the maps taking the dual 
spaces to the spaces we can obtain a map of  the open string
state space to the closed string state space, and a map from the closed string
state space to the open string state space (these maps are not inverses!). Given an
open string state $\ket{\Phi}$, we obtain the closed string state
$$ \ket{\Psi ( \Phi)} = \bra{\V_{oc}} \Phi\rangle \ket{S^c}, \eqn\map$$
where $\ket{S^c} = b_0^- \ket{R_c}$ is the inverse of the closed string 
symplectic form. Since $\ket{S^c}$ is of ghost number five, it follows that 
if the ghost number of $\ket{\Phi}$ is $p$, that of $\ket{\Psi}$ is $p+2$.
Thus we write
$$ \V_{0c} :  {\cal H}_o^{(*)} \rightarrow {\cal H}_c^{(*+2)}\,.\eqn\maptoc$$
In a completely analogous way we also find that
$$ \V_{0c} :  {\cal H}_c^{(*)} \rightarrow {\cal H}_o^{(*)}\,.\eqn\maptop$$
These maps are  dependent on the choices of local coordinates
defining the open-closed string vertex. In other words, they
take different forms for the light-cone and the covariant vertex.
It follows from \xone\ that these maps  induce maps in cohomology;
indeed, BRST trivial open states are mapped to BRST trivial closes string
states (recall that $(Q^{(1)}_c + Q^{(2)}_c ) \ket{S^c_{12}} =0$). We thus have
$$\eqalign{
 \V_{0c} :\,\, \, & H_o^{(*)} \rightarrow H_c^{(*+2)}\cr
 \V_{0c} :\,\, \, &H_c^{(*)} \rightarrow H_o^{(*)}\,.}\eqn\maptcoh$$
The above maps in cohomology are quite interesting and it would be nice
to know more about them. For BRST classes that have primary representatives,
the above maps are independent of  the particular 
choice of open-closed string vertex.  The kernels of the above maps
will play some role in sections 8 and 9. 

\section{The Open-Open-Closed String Vertex}

 In order
to extract this vertex let us consider the scattering of
two open strings and a closed string off a disk. As before, the surface can
be thought as a unit disk with the closed string puncture at the origin. The
two open string punctures can be put symmetrically with respect to the real
axis, as shown in Fig.~13. The angle $\theta$ shown in the figure,
must vary from zero to $\pi$ in order to represent the complete
moduli space. 
In addition to an elementary interaction, the only possible  Feynman graph
for the process in question 
is one built with a three open string vertex and an open-closed vertex,
as shown in Fig.~14.
 
\Figure{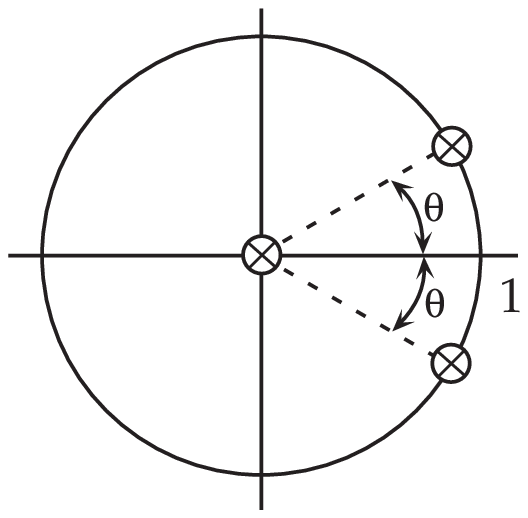 scaled 700}{Figure 13. A disk with two punctures on the boundary and
one in the interior, representing the coupling of two open strings to
a closed string. The open string punctures are placed symmetrically
with respect to the real axis, and the closed string puncture
is placed at the origin. The full moduli space is described
by $0\leq \theta \leq \pi$. }

\Figure{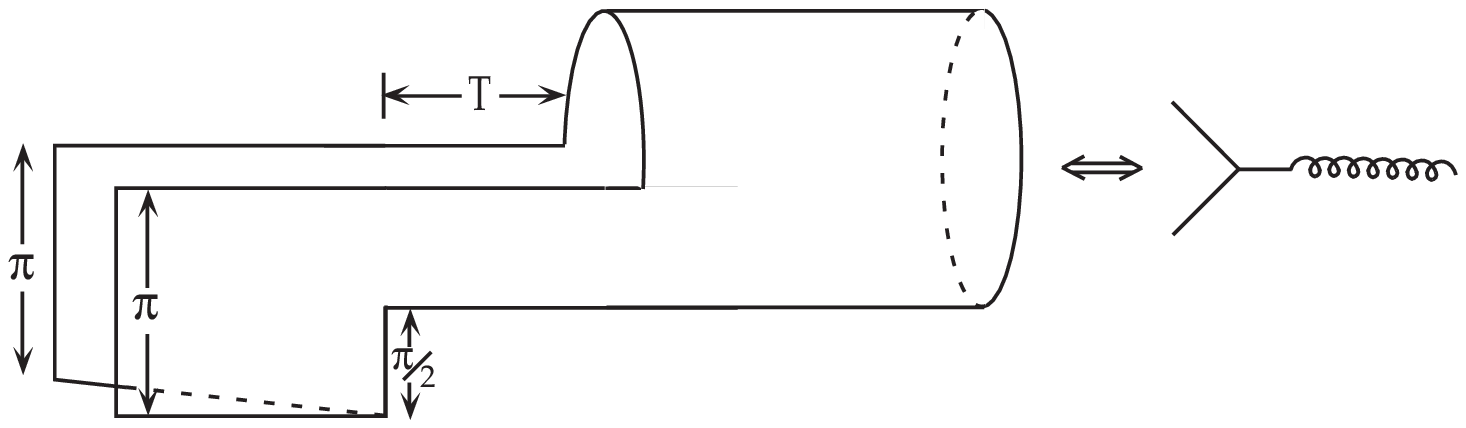 scaled 700}{Figure 14.   A Feynman graph built with a three open
string vertex, an open string propagator of length $T$, and an open-closed vertex. } 

As the open string propagator becomes infinitely
long, namely, $T \to \infty$, the two open string punctures are
getting close to each other, $\theta \to 0$ (Fig.~13). When the
propagator collapses,
$T=0$, the angle $\theta=\theta_0$ is less than $\pi /2$. This is so because 
from the viewpoint of the closed string the open strings are not opposite to
each other. The same Feynman
graph with the punctures exchanged covers the interval
$\pi -\theta_0 \leq \theta \leq \pi$.  The missing region of moduli
space is therefore $\theta_0 < \theta < \pi - \theta_0$. This region
must be generated by the open-open-closed elementary vertex shown
in Fig.~15. Note that the two open strings
ovelap with each other along the segment $AB$, taken
to be of length $a_{12}$. In the endpoint configuration $(T=0)$ for the
Feynman graph of Fig.~14, $a_{12} = \pi/2$. In fact, $a_{12}$ cannot
exceed $\pi /2$;  if it did, the nontrivial open curve $CAD$ shown 
in Fig.~15(a), would be shorter than $\pi$:
$l_{CAD} =l_{CA} + l_{AD} \leq 2 \cdot  \pi/2 = \pi$.
The interaction vertex corresponds to the
region $a_{12} \in [0,\pi/2 ]$. 
The pattern of overlaps is indicated in Fig.~15(b). When $a_{12} = 0$,
the open strings cover the closed string and 
 are opposite to each other. This configuration
corresponds to $\theta = \pi/2$.

\Figure{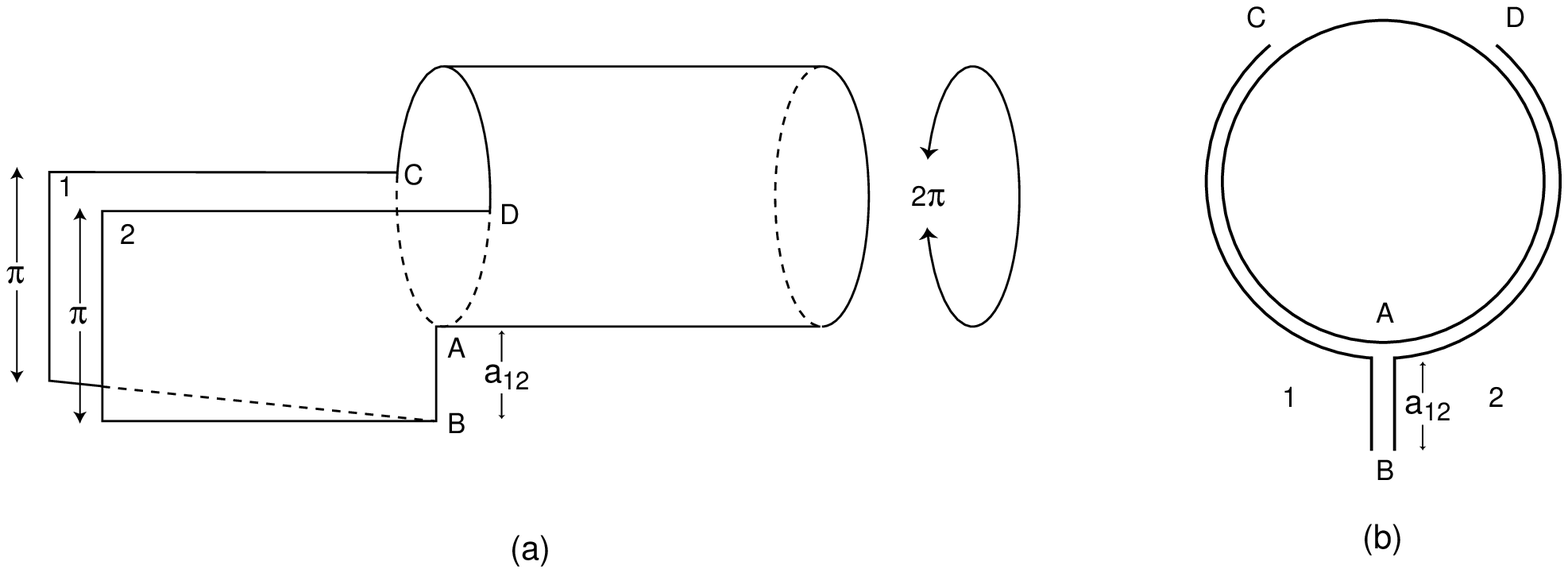 scaled 600}{Figure 15.The interaction vertex for two open strings and
one closed string. The two open strings have an
overlap $AB$ of length $a_{12}$. The rest of the open strings
create part of the emerging closed string. This interaction
has one modular parameter $a_{12}$, which must be smaller
than $\pi /2$, since otherwise the nontrivial open curve $CAD$ 
would be shorter than $\pi$.}

\section{The Vertex Coupling $M$ Open Strings to a Closed String}

Here we  consider the case of $M$ open strings and a single closed string
scattering off a disk.  The  surfaces building the vertex for this process
fall into two different types. In the first type the 
open strings surround completely the closed string. In the second type
the open strings do not surround completely the
closed string; there are $M$ such terms, correponding to the $M$
configurations compatible with the cyclic ordering of the
open strings. 

For both types, the only nontrivial closed curves
are those going around the closed string puncture, and all those
are manifestly longer than $2\pi$. Thus the only relevant constraints
are the ones on open curves. Consider Fig.~16, the condition that
the nontrivial open curve $AA'B'C'C$ be longer or equal to $\pi$
implies that $a_{ij}\leq \pi/2$. In fact, this is the unique
constraint, all open string overlap segments must be shorter or
equal to $\pi/2$. The reader can verify that all other nontrivial
open curves are then automatically longer or equal to $\pi$.

\Figure{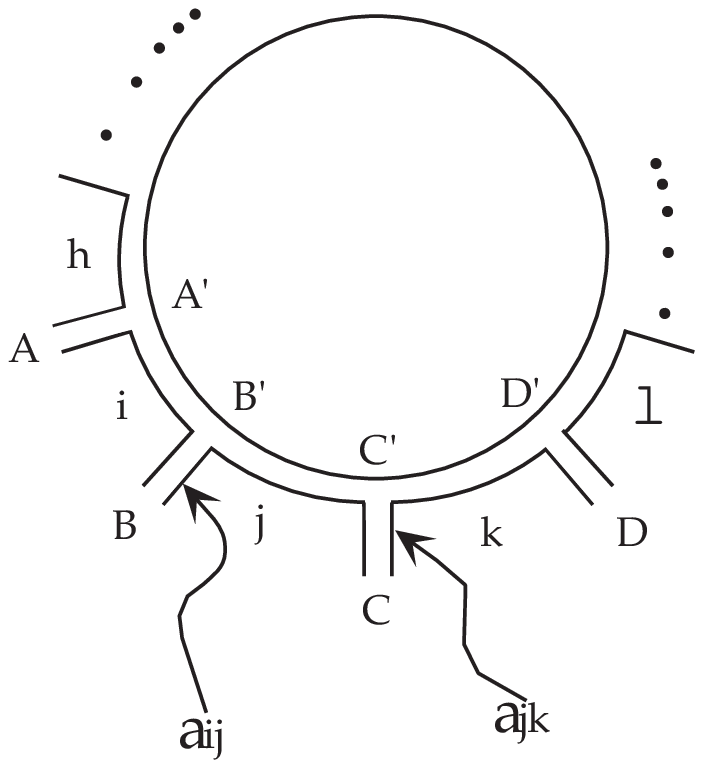 scaled 800}{Figure 16. A generic configuration coupling several open
strings to a single closed string. The necessary and
sufficient condition in order to have all nontrivial open curves
longer or equal to $\pi$ is that all open string overlaps $a_{ij}$
be shorter than or equal to $\pi/2$. }

How about boundary matching? The only inequalities that
restrict the region of integration for the vertex are
$0 \leq a_{ij} \leq \pi/2$.  Boundaries may only
arise when these inequalities are
saturated. Whenever $a_{ij} = 0$, there is actually no boundary.
Consider the configuration where the open strings surround completely
the closed string. Whenever an overlap
segment becomes zero, the configuration turns smoothly into one of the
configurations that do not surround fully the closed string. Whenever
an overlap segment becomes zero in one of the latter configurations
one actually loses all modular parameters of the configuration, so
this is not a relevant boundary. One can understand this by examining Fig.~17(a).
Segments such as $AB$ or $A'B'$ must be at least of length $\pi/2$.
The same is true for segments $CD$ and $C'D'$. These four segments
therefore cover fully the closed string. The only way this is
possible is if all these four segments measure exactly $\pi/2$,
the points $A$ and $A'$ coincide, and every other $a_{ij}$ is
identically equal to $\pi/2$, as shown to the right in the figure.
Indeed,  this configuration has no modular parameters. It also
follows from this argument that there is no relevant configuration where
the set of open strings breaks into two clusters, since each cluster
must at least cover a length $\pi$ in the closed string.
 
Whenever $a_{ij}=\pi/2$ one does get a boundary. Such configuration
matches smoothly with a Feynman graph having strings $i$ and $j$
replaced by a single string $k$ that is connected via a propagator to a three string
vertex having strings $i$ and $j$ as external strings. As illustrated
in Fig.~17(b), when the propagator collapses ($T=0$) the resulting configuration
matches with the vertex boundary. It is clear that for any possible
collapsed propagator configuration there is a corresponding vertex
boundary configuration. Thus the matching of boundaries is complete.
 
Note that the vertex has $(M-1)$ modular parameters. This is most
easily seen from the configuration  where the open strings surround fully
the closed string. In this case the 
$M$ parameters $a_{ij}$ must satisfy a single equality:
$\pi M -2\sum a_{ij} =2\pi$. 
 
\Figure{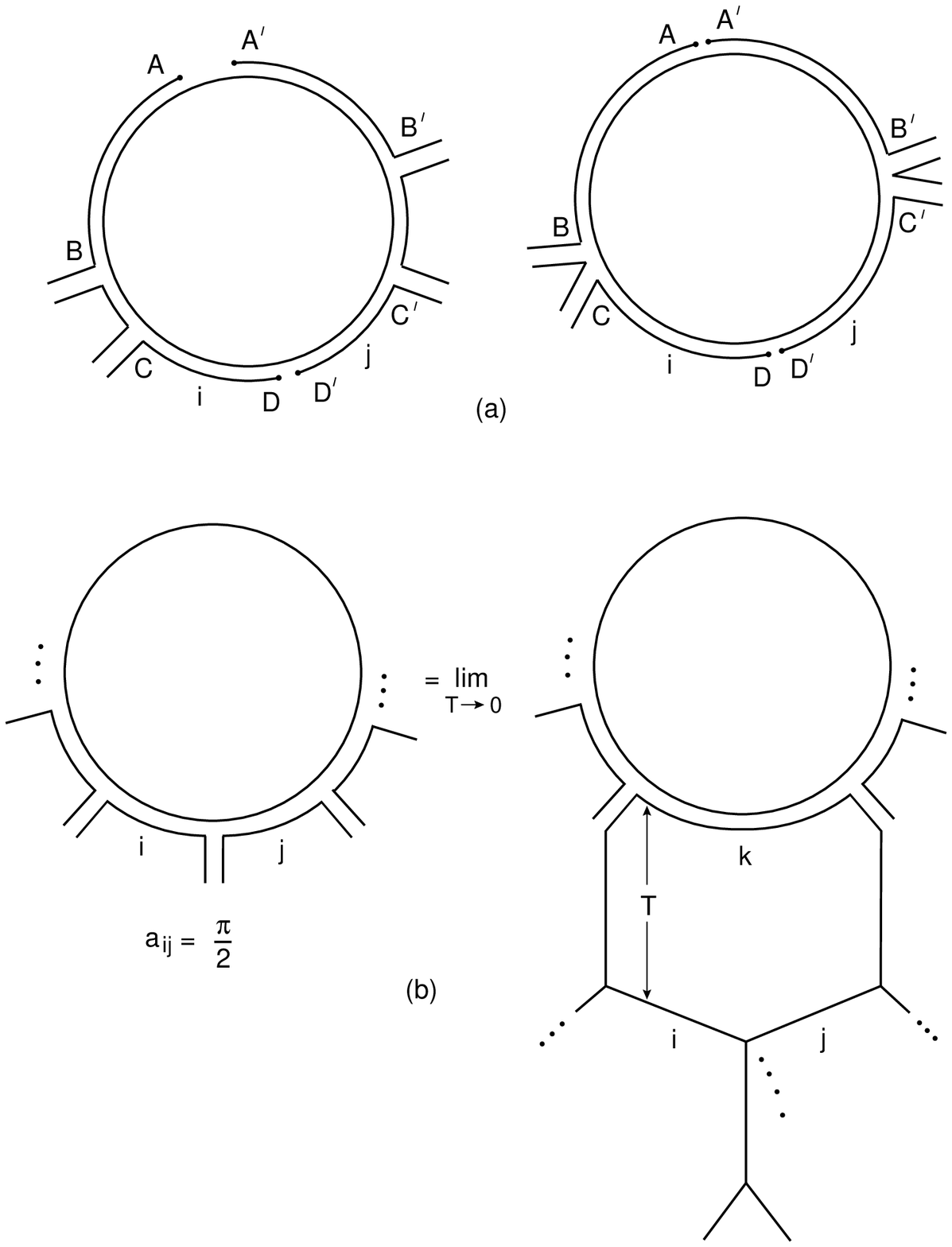 scaled 750}{Figure 17. (a) A configuration where an overlap $a_{ij}$
vanishes,
 is not a codimension one
boundary. It can be shown that it has no modular
parameters, and must really look like the diagram to the right.
(b) When some $a_{ij}$ becomes $\pi/2$, the vertex shown to the
left matches with the $T=0$ limit of the Feynman graph
shown to the right. This Feynman graph uses a vertex coupling
$(M-1)$ open strings to a closed string, and a three open string vertex.}

\Figure{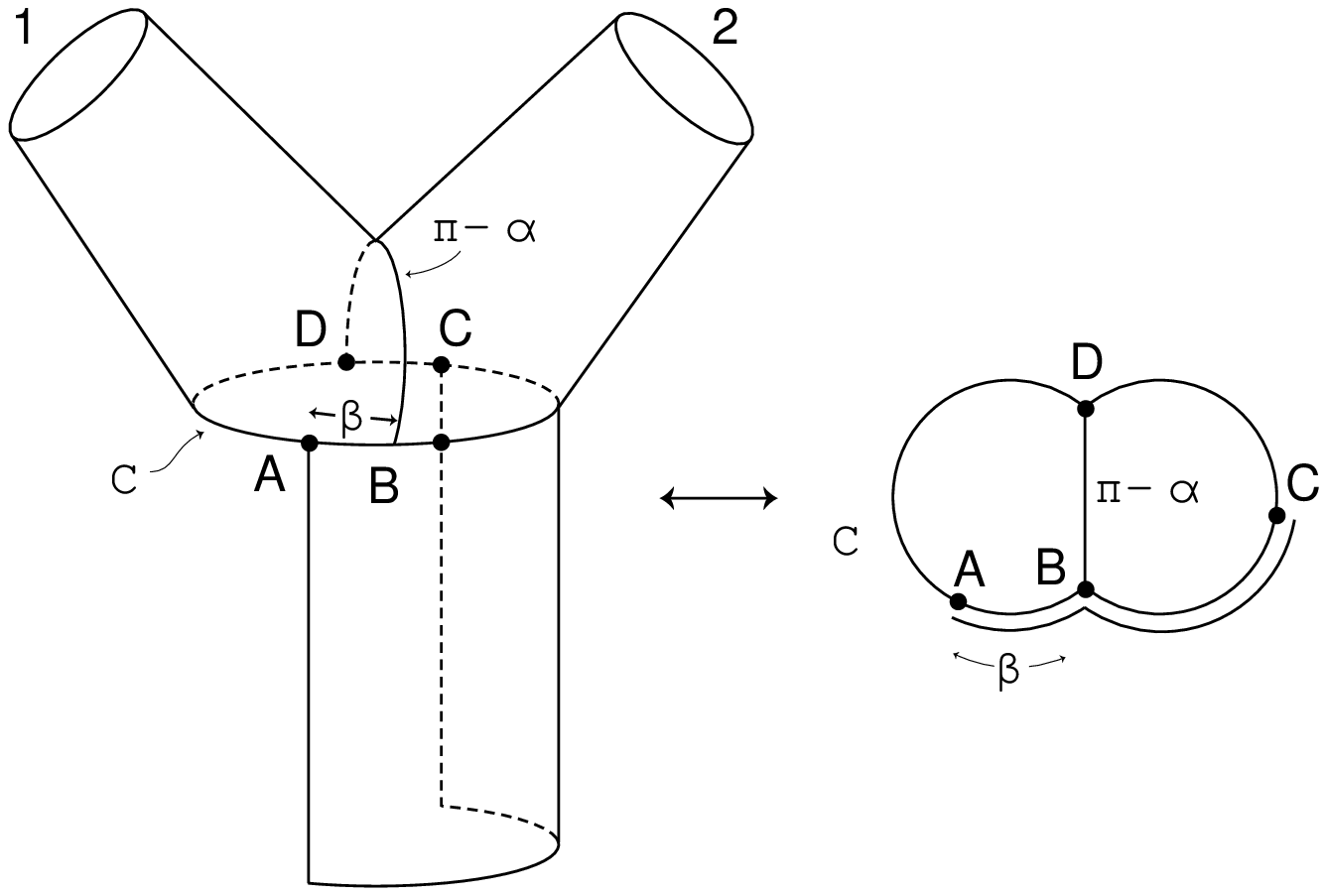 scaled 900}{Figure 18. The covariant vertex for closed-closed-open
strings. The two closed strings overlap across the arc $BD$ of length
$(\pi -\alpha)$, with $\alpha \geq 0$. The closed curve $\C$
going through $A,B,C,D,$ and back to $A$, is longer than $2\pi$.
The open string emerges to the bottom of the diagram. 
To the right we show the pattern of overlaps in this interaction. }

\section{The Open-Closed-Closed Vertex} 

The vertices
coupling two closed strings to open strings via a disk are relevant
for the elucidation of the  algebra of global symmetries
induced by closed strings. We discuss next the
simplest such vertex, having two closed strings and a single open
string. This vertex fills a region of moduli space of dimension
two, and the generic surface is shown in Fig.~18. The two closed
strings, labeled $1$ and $2$,  overlap across the arc $BD$, and create a closed boundary
$\C$, part of which is attached to the open string. The corresponding
overlap pattern is shown to the right. Note that the arc $BD$ must
be shorter than $\pi$, otherwise the closed curve $\C$ would be
shorter than $2\pi$. The length of $BD$ is thus $(\pi -\alpha)$,
with $\alpha \geq 0$. The curve $\C$ has length greater or equal
to $2\pi$. The open string must overlap with both closed strings,
as shown in the figure, if it did not, the arc $BD$, which is
shorter than $\pi$, would become a nontrivial open arc, thus
violating a length condition. The length of the segment $AB$,
where the open string and closed string number one overlap, is
denoted by  $\beta \geq 0$. 

There are two open paths whose length must be checked:
$$l_{ABD} = l_{AB} + l_{BD} = \beta +(\pi-\alpha) \geq \pi\quad
\to \quad\alpha \leq \beta,\eqn\yone$$
$$l_{CBD} = l_{CB} + l_{BD} = (\pi-\beta) + (\pi-\alpha )\geq \pi\quad
\to \alpha + \beta\leq \pi .\eqn\ytwo$$
These two inequalities, together with $\alpha, \beta  \geq 0$, make the region
indicated in Fig.~19(a). This is actually half the moduli space, the
other half corresponds to the open string glued to $\C$ such that
point $B$ is on the boundary, and point $D$ is not.

What do the boundaries of the region in Fig.~19(a) match to?
Consider first the boundary $\alpha = 0$. In this configuration 
the two closed strings are overlapping
like in the three closed string vertex. Indeed, the length of $\C$ is
precisely $2\pi$. This boundary matches with a Feynman graph
where we have a three closed string vertex  joining to an closed-open
vertex via a closed string propagator, in the limit when this closed
string propagator collapses and becomes the curve $\C$. 

The other two boundaries appear from the same type of process:  the
appearance of an intermediate open string propagator. They correspond
to the nontrivial open curves $ABD$ and $CBD$ considered above
becoming of length $\pi$.  When $l_{ABD} = \pi$ an open strip can 
grow  separating closed string $1$ from the rest of the interaction. The
Feynman graph it corresponds to is one where a open-closed vertex turning
closed string $1$ into an open string is joined by an open string propagator
to an open-open-closed vertex.  As indicated in Fig.~19(b),  
the open string to the left closes up to make closed string number one. 
This configuration is recognized to have $\alpha = \beta$. Analogous remarks apply to
$l_{CBD}$ becoming of length $\pi$ (Fig.~19(c)).   

\Figure{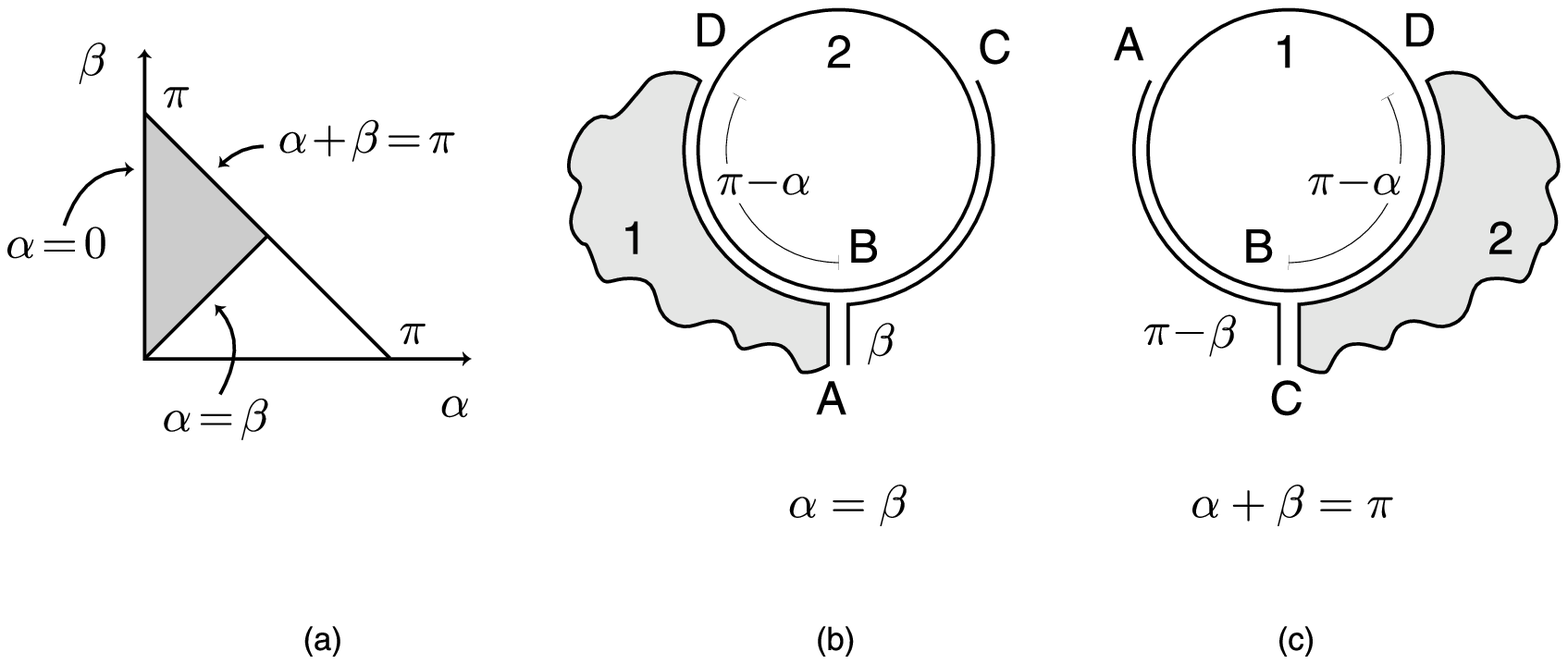 scaled 800}{Figure 19. (a) The parameter region for $\alpha$ and
$\beta$ in the closed-closed-open vertex shown in Fig.~12. There are three
boundaries. When $\alpha = 0$ the nontrivial closed
curve $\C$ becomes of length $2\pi$. (b) This configuration 
matches the $\alpha = \beta$ boundary of the parameter region. It
corresponds to the nontrivial open curve $ABD$ becoming of length $\pi$.
(c) This configuration 
matches the $\alpha +\beta=\pi$ boundary of the parameter region. It
corresponds to the nontrivial open curve $CBD$ becoming of length $\pi$. } 

\section{Vertex Coupling $M$ Open Strings to two Closed Strings}

The geometrical arrangement for this coupling follows from the
previous case together with that of the coupling of many open
strings to a single closed string. As in the previous case, the
two closed strings join via an arc of length less than $\pi$, and
form a closed curve $\C$, as was the case in Fig.~18. The $M$ open strings
then couple as if $\C$ was a closed string. There is just one qualitative
difference; since the `closed string' $\C$ is longer than $2\pi$ the
cluster of $M$ open strings can break into subclusters. In the present
case, since the maximum possible length of $\C$ is $4\pi$, the open strings 
can break nontrivially  into two or three separate clusters, each of which must cover at
least a length $\pi$ on $\C$. 
One must still require that all overlaps $a_{ij}$ between open
strings be  less than  or equal to $\pi/2$. There are additional constraints, however, corresponding
to nontrivial open curves that use the arc joining the two closed strings
 
The counting of modular parameters is simple if one refers to the
case when the open strings surround $\C$ completely. 
There are $M$ parameters for the open string overlaps,
one for the length of $\C$, and one constraint relating them, giving
$M$ parameters. There is one extra parameter coming from a global
rotation of the cluster of open strings along $\C$. This
gives a total of $M+1$ modular parameters. 

The boundaries are of two types, boundaries matching to Feynman graphs
with a closed string propagator and boundaries matching to Feynman graphs
with an open string propagator. The first type arises when 
the closed curve $\C$ becomes of length
$2\pi$; the Feynman graph is built from a three closed string vertex and a 
vertex coupling one closed string to $M$ open strings. The second 
type falls into two cases. In the first one an overlap $a_{ij}$ of
open strings becomes of length $\pi/2$, and  we get an open string propagator 
separating away open strings $i$ and $j$ (see Fig.~17(b)). 
In the second one, a nontrivial open curve  that includes the arc separating the
two closed strings, becomes of length $\pi$. Here an open string propagator
can grow separating out  one of the closed strings and some number of open strings.

\section{Vertices Coupling $N$ Closed Strings to $M$ Open Strings}

This is the most general coupling via a disk. Thanks to the
practice gained in the previous examples it can be dealt with
easily. The $N$ closed strings make up a polyhedron with
$(N+1)$ faces. Each of the $N$ faces corresponding to the closed
strings is of perimeter $2\pi$. The remaining face is left open,
we call its boundary $\C$, and it is a closed curve of length
greater or equal to $2\pi$. All closed paths in this polyhedron
must be longer or equal to $2\pi$. The number of modular parameters
of this polyhedron is one more than that of a standard $(N+1)$-faced
polyhedron, because the open face has no perimeter constraint.
Since the latter has $2(N+1)-6$ parameters, this gives
$2N-3$ modular parameters for the open-faced polyhedron.
 
The cluster of $M$ open strings is attached to the closed curve
$\C$, just as in the previous case. This cluster can again break
into several clusters, depending on the length of $\C$. For
fixed length of $C$, the open strings add $(M-1)+1$ parameters,
where the last one comes from twist. The total number of
modular parameters is therefore $2N+M-3 = 2(N-1)+(M-1)$.

There are three types of nontrivial closed curves on the
polyhedron: (a) closed curves that
do not use any arc of $\C$, (b) closed curves that use pieces of $\C$,
and (c) the closed curve $\C$.  When any of these curves becomes
of length $2\pi$ we match to a Feynman graph with an intermediate
closed string propagator.  Both for cases (a) and (b) the diagram
breaks into a pure closed string vertex coupling  a subset of the $N$
external closed strings, and a vertex coupling the remaining closed
strings to all of the $M$ open strings. For case (c) the diagram breaks
into a pure $N+1$ closed string vertex and a vertex coupling one
closed string to all of the open strings. Note that  for all closed string intermediate
channels  the open strings must remain together since otherwise one
would get two boundary components.

 There are two types of nontrivial open curves: (i) curves that do not use
arcs joining closed strings, and (ii) curves that do.  The first type of curves
saturate when an overlap $a_{ij}$ becomes of length $\pi/2$, 
and  we get an open string propagator 
separating away open strings $i$ and $j$ (see Fig.~17(b)). 
When type (ii) curves saturate, an open string intermediate strip
separates the vertex into two vertices each carrying a nonzero
number of closed strings and some number of external open strings.

\chapter{Open string symmetries of closed string origin}

In this section we discuss symmetries of classical open string theory that
arise from the closed string sector. Such symmetries were first discussed
by Hata and Nojiri [\hatanojiri] in the context of covariantized light-cone
string field theories. Due to the much improved understanding of 
the recursion relations of string vertices and of BV quantization of 
string theory, our discussion will be quite transparent. We will also
be able to discuss the relation between these trasformations and
the usual open string gauge transformations, as well as elucidating the
algebra of such symmetries.

\section{Symmetry transformations}

We are dealing with symmetries of the 
classical open string action. We recall that this action can be written as
$$S_o = Q_o + f(\V_0)\,,\eqn\wthree$$
where $\V_0$ is the formal sum of all the string vertices 
coupling open strings only through a disk. The consistency of
this action follows from the recursion relation \rectree\ which reads
$$ \partial \V_0 + \half \{ \V_0 \,, \, \V_0 \} = 0\,. \eqn\rectree$$
We now consider the symmetry transformation generated by the hamiltonian
$$U = f_\O ( \V_1)\,, \eqn\stcso$$
where $\O$ is an arbitrary  closed string state {\it annihilated } by $Q_c$, and
$\V_1$ is the sum of string vertices  coupling all numbers of 
open strings to a single closed string through a disk.
We recall that these vertices satisfy the recursion relation \recclo\ which
reads
$$\partial \V_1 +  \{ \V_0 \,, \, \V_1 \} = 0 \eqn\recclo$$
The variation of the action under an infinitesimal transformation
generated by $U$ is given by
$$\eqalign{
\delta S_o =  \{ S_o , U \} &=  
\Bigl\{ Q_o + f(\V_0)\, , f_\O (\V_1) \Bigr\}\cr
& = -f_\O \Bigl( \partial \V_1 +  \{ \V_0 \,, \, \V_1 \} \Bigr)
 - f_{Q_c\O}(\V_1) \,  \cr}  \eqn\provesymm$$
where we added and subtracted a term involving the closed string BRST
operator in order to produce the boundary action on $\V_1$. It is
now evident that $\delta S_o$ vanishes 
by virtue of the recursion relation  \recclo\ and the condition
$Q_c\ket{\O} =0$. This establishes that 
\stcso\ generates a symmetry. The open string field variation induced
by this hamiltonian is simply
$$\delta \ket{\Phi} = \{ \ket{\Phi} , U \} \sim 
{\partial\, U \over \partial \,\ket{\Phi}} \,\ket{S^o}\,, \eqn\howl$$ 
and given that $\,U \sim \bra{\V_{oc}} \O\rangle \ket{\Phi} + \cdots\,,$
we have that
$$\delta \ket{\Phi}\sim \bra{\V_{oc}} \O\rangle \ket{S^o} + \cdots \,. \eqn\efor$$
The symmetry transformation, to first approximation, shifts the open
string field along the image of the closed string state $\ket{\O}$ under
the map \maptop\ defined by $\V_{oc}$. The nonlinear terms of the
transformation involve one or more open string fields and use
the vertices  coupling several open strings to a single
closed string.

Since the open closed vertex maps closed string states of a given
ghost number to open string states of the same ghost number (see \maptop)
 BRST closed states of ghost number one in
the closed string sector are of particular interest. 
Such states would induce nontrivial transformations
in the physical ghost number one sector of the open string field.  Indeed
we know that closed string BRST classes at ghost number one correspond to
global symmetries of the closed string sector. This point has been made
very precise by defining the proper BRST cohomology taking into 
account zero modes of the bosonic coordinates [\astashkevich,\belopolskyzw].
In the bosonic context Poincare transformations arise from the closed
string  BRST cohomology
at ghost number one. The corresponding 
open string transformation defines the induced Poincare transformation of
the open string field.

We wish to examine  what happens when the closed string state $\ket{\O}$
used to define the open string transformation is $Q_c$-trivial. We will show
that in this case we simply get a gauge transformation of the open string
sector.  This confirms that only nontrivial $Q_c$  classes generate new
transformations. Recall that open string gauge transformations are generated by
Hamiltonians $U$ of the form
$$U = \{ S_o , \Lambda \} \,.\eqn\osgt$$
Choose now $\Lambda = f_\eta (\V_1)$ where $\eta$ is a closed string state.
We then find
$$\eqalign{
U &= \{ S_o , f_\eta (\V_1) \}\cr
&= \{Q_o + f(\V_0) \, ,  f_\eta (\V_1)\} \cr
&= -f_{Q_c\eta} (\V_1) - f_\eta \Bigl(
\partial\V_1 + \{ \V_0\,,\, \V_1\}\Bigr) \cr
& = -f_{Q_c\eta} (\V_1) \,. \cr}\eqn\osgt$$
Comparing the last and first lines of the above equation we see that 
indeed, an open string transformation induced by $Q_c\eta$ can be written 
as an open string gauge transformation induced by the gauge parameter
$f_\eta(\V_1)$.

\section{ The Algebra of Symmetry Transformations}

Let us now examine the algebra of closed string induced open-string 
transformations. The result is simple: the commutator of two such
transformations gives another closed string induced transformation,
and an open string gauge transformation.  To show this consider two
$Q_c$-closed string states $\ket{\O_1}$ and $\ket{\O_2}$ and
the associated hamiltonians $U_1 = f_{\O_1}(\V_1)$ and 
$U_2 = f_{\O_2}(\V_1)$. The hamiltonian defining the commutator
of two such transformations is simply the antibracket of the two
hamiltonians. We find
$$\{  U_1\,, \, U_2 \}
=\{  f_{\O_1}(\V_1)\,, \,  f_{\O_2}(\V_1) \} =
-f_{\O_1\O_2} \Bigl( \{ \V_1 \,,\, \V_1 \}_o \Bigr) \,,  \eqn\twocom$$
where the subscript in the antibracket reminds us that sewing is only
performed in the open string sector (this must be the case since we
cannot lose the two closed string punctures). We now recall the recursion
relation \rectwo\ involving the open-closed vertices $\V_2$ having two closed
string insertions
$$\partial \V_2 + \{\V_0, \V_2\} + \half \{\V_1 , \V_1 \}_o + 
\{\V_1, \V_3^c\}= 0 \,. \eqn\rectwo$$
Back in \twocom\ we get
$$\eqalign{
\{  U_1\,, \, U_2 \} &= 2 f_{\O_1\O_2}\Bigl(\partial \V_2 + \{\V_0, \V_2\}
\Bigr) +  2 f_{\O_1\O_2}\Bigl( \{\V_1, \V_3^c\} \Bigr)\cr 
&= - 2\, \Bigl\{ S_o \,,\,   f_{\O_1\O_2}(\V_2) \Bigr\}
 -  2 \,\Bigl\{  f(\V_1)\, ,\, f_{\O_1\O_2}(\V_3^c) \Bigr\}\,, \cr} 
\eqn\backcom$$
where we made use of the fact that the states $\ket{\O_1}$ and $\ket{\O_2}$
are $Q_c$ closed. The first term in the last right hand side is 
recognized as a hamiltonian inducing an open string gauge transformation.
The second term is recognized as being of the form $f_\Omega (\V_1)$ where
the closed string state $\Omega$ is the state induced by inserting 
$\ket{\O_1}$ and $\ket{\O_2}$ on two of the punctures of the three closed
string vertex $\V_3^c$. More precisely $\ket{\Omega} = \bra{\V_3^c}\O_1\rangle
\ket{\O_2} \ket{S}$. This confirms the claim for the commutator of two
closed string-induced open string symmetries. 

The commutator of a closed string-induced symmetry transformation and
a open string gauge transformation is an open string gauge transformation.
This is easily verified, 
$$\Bigl\{ f_\O (\V_1)\,, \{ S_o \,, \lambda\} \Bigr\} \sim 
\Bigl\{  S_o\,, \{ \lambda\,, f_\O (\V_1) \} \Bigr\} +
\Bigl\{ \lambda\,, \{ f_\O (\V_1)\,,  S_o\} \Bigr\}\,, \eqn\area$$
by use of the Jacobi identity of the antibracket. The last term in
the right hand side vanishes (\provesymm) and the first term indeed
represents the hamiltonian for an open string gauge transformation.

\chapter{Open string theory on closed string backgrounds}

This time we are interested in a truncation of the full open-closed
theory to a classical open string sector in interaction with an
explicit closed string background. The classical open string field theory
of \wthree\ is one such system, and indeed it is classically gauge 
invariant. This action is obtained by setting to zero the fluctuating
closed string field in the open-closed string theory action. It thus
represents open string theory on the closed string background described
by the conformal theory having $Q_c$ as the BRST operator. 
We now imagine giving the closed string field an 
expectation value $\ket{\Psi_0}$ thus changing the closed string
background in a controlled way. What we will show is that we can
obtain a consistent classical open string theory whenever $\ket{\Psi_0}$
satisfies the closed string field equation of motion. This action
represents open string theory on the closed string background 
described by $\ket{\Psi_0}$.

We claim that the open-string action is given as
$$S_o (\Phi, \Psi) = 
Q_o + \sum_{k=0}^\infty \V_k \equiv Q_o + f(\ov\V) \,,\eqn\claimi$$
where we have included now all couplings of open and closed strings
through disks. This action depends both on the open string field
and the closed string field, but we will think of the closed string field
as non-dynamical. For consistency this action should satisfy the
classical master equation. We compute
$$\{S_o \,,\, S_o \}_o = 2 \{ Q_o \,, f(\ov\V) \} 
+ \{ f(\ov\V)\,,\, f(\ov\V)\}_o\,, \eqn\oscsb$$
where the antibracket has been restricted to the open string sector,
given that the closed string field is not dynamical. Now 
add and subtract  a closed string BRST operator to find
$$\eqalign{
\{S_o \,,\, S_o \} &= 2 \{ Q_o + Q_c \,, f(\ov\V) \} 
+ \{ f(\ov\V)\,,\, f(\ov\V)\}_o -  2 \{ Q_c \,, f(\ov\V) \} \cr
 &= -2 f \Bigl( \partial\ov\V
+ \half\{ \ov\V\,,\, \ov\V\}_o \Bigr) -  
2 \{ Q_c \,, f(\ov\V) \}\,,\cr} \eqn\oscsbb$$
In order to proceed we must recall the  recursion relation \reopct\ satisfied
by the $\ov\V$ spaces
$$\partial \ov\V + \half \{ \ov\V \,,\, \ov\V \}_o + \{ \ov\V\,,\,\V^c \} 
=0\,, \eqn\reopc$$
where $\V^c$ is the sum of pure closed string vertices on the sphere.
Back in \oscsbb\ we obtain 
$$\eqalign{
\{S_o \,,\, S_o \}
 &= 2\, f\, ( \{ \V^c\,,\, \ov\V\} ) -  
2 \{ Q_c \,, f(\ov\V) \}\,,\cr
&= - 2 f\Bigl( \{ Q_c + f(\V^c) \,,\,  f(\ov\V) \,\}\Bigr) \,,\cr
&= - 2 f\,( \{ S_c \,,\,  f(\ov\V) \,\}\,) \,,\cr} \eqn\csbb$$
where in the last step we have recognized the classical closed string
action. Since
$$ \{ S_c \,,\,  f(\ov\V) \,\} \sim  
{\partial\, S_c\over \,\partial \ket{\Psi}} 
\,{\partial\, f(\ov\V) \over \partial\, \ket{\Psi}} \, \ket{S} \,,\eqn\alms$$
this term vanishes for closed string fields satisfying the closed string
field equation of motion. In this case we find $\{S_o, S_o \}_o =0$ as
we wanted to show. This verifies the consistency of this generalized
classical open string field theory.  As usual, the master equation guarantees
the gauge invariance of the theory, and we have that hamiltonians of the
type $U= \{S_o, \Lambda\}_o$ generate gauge transformations. Such gauge 
transformations depend on the closed string background, since $S_o$ 
depends on the closed string background. 

The reverse construction, trying to describe a closed string sector
moving on an open string background, does not seem possible. 
The obvious ansatz $S_c = Q_c + f(\V_c) + f(\overline\V)$ fails.
Attempts made using other subalgebras of surfaces, such as that
of all genus zero surfaces (see \nouse) failed as well.

\chapter{On the ghost-dilaton theorem and background independence of open-closed theory}

In this section we sketch briefly the main issues that arise when one 
attempts to establish a ghost-dilaton theorem in open-closed string theory. 
We also comment on the similar issues that arise in trying to 
prove background independence.

It is useful, for clarity, to write out explicitly
the first few terms of the open-closed string action.
To this end we use the list of vertices apppearing at the
first few orders of $\hbar$ as given in section 3.2.  We have, without attention to 
dimensionless coefficients, 
$$\eqalign{
S &\sim \bigl( \langle \Phi ,Q_o \Phi \rangle_o + 
\kappa \Phi^3 + \kappa^2 \Phi^4 + \cdots \bigr)
+ \langle \Psi ,Q_c \Psi \rangle_c  \cr
& \quad\, + \hbar^{1\over 2}   \bigl(  \kappa^2 \Psi^3_{sph} +  \Psi_{disk} 
+ \kappa (\Psi\, \Phi)_{disk} 
+ \kappa^2 (\Psi\, \Phi^2)_{disk} + \kappa^3 (\Psi\, \Phi^3)_{disk} \cdots \bigr)  \cr
& \quad\, + \hbar \bigl(  \kappa^4 \Psi^4_{sph} + Z_{torus}  +   \kappa^2 \Psi^2_{disk} + 
\kappa^3 (\Psi^2\Phi)_{disk} + \cdots + Z_{ann} + \kappa\Phi_{ann} + \cdots \bigr) \cr
& \quad\, + \hbar^{3\over 2} \bigl(  \kappa^6 \Psi^5_{sph}  +  
\kappa^2 \Psi_{torus}   + \kappa^4 \Psi_{disk}^3 +  
\kappa^5 (\Psi^3 \Phi)_{disk}  +  \cdots 
\bigr)  \cr
&\quad + \cdots \cr}  \eqn\actionexp$$
Here the subscripts indicate the type of surface involved in the vertex, 
with $sph$ standing
for sphere and $ann$ standing for annulus.

In order to  establish a ghost dilaton theorem, as that
proven for closed strings in Refs.[\bergmanzwiebach,\rahmanzwiebach], we 
have to show that there is a dilaton hamiltonian $U_D$ that generates a string field
redefinition that changes the string coupling in the action. In particular, at the full
quantum level, the dilaton hamiltonian must satisfy
$$\kappa {d S\over d\kappa} = \{ S , U_D\} + \hbar \Delta U_D \,, \eqn\needprove$$
as  discussed in section 4 of Ref.[\bergmanzwiebach]. The dilaton hamiltonian $U_D$
must begin with a term of the type $\langle D, \Psi\rangle_c$ 
 that gives a shift in the closed string field along the dilaton direction $\ket{D}$. The shift
must be of the form
$$\ket{\Psi} \to   {1\over \hbar^{1/2} \kappa^2} \ket{D} +  \ket{\Psi} \, ,\eqn\nhg$$
where we have inserted the proper factors of $\hbar$ and string coupling $\kappa$
necessary for the shift to have the correct effect. Indeed, we 
see in \actionexp\ that each  closed string insertion contributes
a prefactor of $\hbar^{1/2}\kappa^2$ to the vertices. This shift will not only
change the string coupling as it appears in purely closed string terms, but also
ought to change the string coupling in terms involving open strings.  It  certainly
has the potential to do so; for example, the term $\kappa (\Phi^3)_{disk}$ can have its
coupling changed due to the term $\hbar^{1/2}\kappa^3 (\Phi^3 \Psi)_{disk}$.  
There are a few terms,
however, that introduce interesting complications.

 The coupling  $\langle \Psi, B\rangle_c$ 
of the closed string to the boundary state (section 6.1) indicated above as
$\Psi_{disk}$, and  appearing at order $\hbar^{1/2}$ with
no coupling constant dependence is one of those terms. When we shift by the ghost dilaton,
we get a constant term of the form
$$  {1\over \kappa^2 } \hbar^0 \langle D, B \rangle_c \,, \eqn\constf$$
This constant has the dependence to be interpreted as arising from a disk with no
punctures (which is precisely what we get when the closed string puncture is erased).
This implies that we must include in the above action a {\it tree level cosmological
term} arising as an open string partition function on the disk
$$ S=  {1\over \kappa^2}\,   Z_{disk} + \cdots \,\, , \eqn\getcosm$$
since otherwise \nhg\ could not be satisfied. \foot{In early discussions of these 
issues in the framework of effective actions,   terms that could yield cosmological
constants were attributed to open string loop corrections to closed string beta
functions [\callan].}Note that the effect of shifting along 
the dilaton producing \constf\ is due to
 the first term in the right hand side of \nhg.  The second
term in \nhg\  can also produce a constant, but its geometrical interpretation
indicates that it  refers to surfaces of higher genus or surfaces with 
more than one boundary component. 
The overlap  $\langle D, B \rangle_c$ of the ghost-dilaton with the
boundary state does not vanish in general. In fact, it does not vanish for any
oriented open string theory. It can vanish in non-oriented open string theories where
there is a similar contribution arising from a closed string ending on a crosscap. 
This cancellation only happens when
the gauge group of the open string is $SO(8192)$ [\sagnotti].  As we see here, only 
for this gauge group we expect the tree level cosmological constant to vanish.

A second term that raises some puzzles is the term  
$\hbar^{1/2} \kappa (\Psi \Phi)_{disk}$,
involving the open-closed vertex $\bra{\V_{oc}}$.  When the 
closed string field is shifted by the dilaton we now get a term
linear  in the open string field
$$ {1\over \kappa} \bra{\V_{oc}} D\rangle \ket{\Phi} \,, \eqn\linosf$$ 
Since such a term cannot exist in the resulting open-closed string action, it must
be cancelled by giving the open string field a vacuum expectation value
$$\ket{\Phi} \to \ket{\Phi} + {1\over \kappa }\ket{d}\, . \eqn\osvev$$
This shift of the kinetic term produces a linear term that cancels \linosf\
provided that  
$$ Q_o \ket{d}  \sim \bra{\V_{oc}} D\rangle \ket{S^o}\, . \eqn\hutpe$$
The right hand side of this equation is simply the image of the ghost dilaton
under the open-closed map (section 6.2). This equation can  be solved for $\ket{d}$
only if the open-closed map takes the cohomology class of $\ket{D}$ to zero, or in other
words, to a trivial open string state.  This is the case, as can be verified by using
$\ket{D} = Q_c \ket{\chi}$, where $\ket{\chi} = c_0^- \ket{0}$.\foot{Note that
the ghost-dilaton $\ket{D}$ is not trivial because  $\ket{\chi} = c_0^-\ket{0}$
is not a state in the closed string complex. We can nevertheless use this 
equation here because $\V_{oc}$ is well defined in the complete state space
$\H_{cft}$.}  One readily
finds that $\ket{d} \sim \bra{\V_{oc}} \chi\rangle \ket{S^o}$ and therefore the open string 
shift is of the form
$$\ket{\Phi} \to \ket{\Phi} + {1\over \kappa }\, 
\bra{\V_{oc}} \chi\rangle \ket{S^o}\,, \eqn\fds$$
a shift along an unphysical direction. 

Rather similar issues arise in a background independence analysis.
If we shift along a closed string direction specified by a dimension
zero primary state $V$ of ghost number two, we may also generate a 
constant term from the closed-boundary interaction. This will have
the interpretation of a change in the partition function of the disk.
Moreover, a linear term in the open string field can arise. There should
be no difficulty in removing it by an unphysical shift in the open
string. This follows because the open closed map would map closed string
cohomology at ghost number two to open string cohomology at ghost number
two, and such cohomology is known to vanish in the appropriate
complex where we include the zero modes of the non-compact bosonic
coordinates [\astashkevich].

\medskip
\bigskip
I am grateful to Rachel Cohen and Marty Stock for preparing the many figures
contained in this paper.

\refout
\end